\documentclass[journal,10pt]{IEEEtran}
\usepackage[T1]{fontenc}
\usepackage{cite}

\usepackage{fancyhdr}
\usepackage{lastpage}
\usepackage{lipsum} 
\usepackage{amsthm}
\usepackage{amssymb}
\usepackage{amsfonts}
\usepackage{fancyhdr}  %
\usepackage{extarrows}
\usepackage{algorithm}
\usepackage{multirow,tabularx}
\usepackage{mathtools}
\usepackage{xcolor}
\usepackage[english]{babel}
\usepackage{caption}
\usepackage{bm}
\usepackage{algorithmic}
\usepackage{amsmath}
\usepackage{subfigure}
\usepackage{makecell}
\usepackage{colortbl}
\usepackage{booktabs}
\usepackage{amsmath}
\usepackage{diagbox}
\usepackage{makecell}
\usepackage{cancel}
\usepackage{tablefootnote}
\usepackage{balance}
\usepackage{flushend}
\usepackage{afterpage}

\ifCLASSINFOpdf
  \usepackage[pdftex]{graphicx}
  \graphicspath{{../pdf/}{../jpeg/}}
  \DeclareGraphicsExtensions{.pdf,.jpeg,.png}
\else
  \usepackage[dvips]{graphicx}
  \graphicspath{{../eps/}}
  \DeclareGraphicsExtensions{.eps}
\fi
\begin{document}
\title{
Channel Deduction: \\A New Learning Framework to Acquire Channel from Outdated Samples and Coarse Estimate
}
\author{\IEEEauthorblockN{Zirui~Chen,~\IEEEmembership{Graduate~Student~Member,~IEEE, }
Zhaoyang~Zhang,~\IEEEmembership{Senior~Member,~IEEE, } \\
            Zhaohui~Yang,~\IEEEmembership{Member,~IEEE, }
            Chongwen~Huang,~\IEEEmembership{Member,~IEEE, }
            and Mérouane~Debbah,~\IEEEmembership{Fellow,~IEEE } }          
\thanks{This work was supported in part by National Natural Science Foundation of China under Grants 62394292 and U20A20158, Ministry of Industry and Information Technology under Grant TC220H07E, Zhejiang Provincial Key R\&D Program under Grant 2023C01021, and the Fundamental Research Funds for the Central Universities No. 226-2024-00069. (\textit{Corresponding Author: Zhaoyang~Zhang})}
\thanks{Z.~Chen, Z.~Zhang,  Z.~Yang and C.~Huang  are with College of Information Science and Electronic Engineering, Zhejiang University, Hangzhou 310027, China, and also with Zhejiang Provincial Key Laboratory of Info. Proc., Commun. \& Netw. (IPCAN), Hangzhou 310007, China. (E-mails: \{ziruichen, ning\_ming,  yang\_zhaohui, chongwenhuang\}@zju.edu.cn)}
\thanks{M.~Debbah is with Khalifa University of Science and Technology, P O Box 127788, Abu Dhabi, UAE. (E-mail: merouane.debbah@ku.ac.ae)}
}

\maketitle 
\thispagestyle{empty}
\begin{abstract}
How to reduce the pilot overhead required for channel estimation? How to deal with the channel dynamic changes and error propagation in channel prediction? To jointly address these two critical issues in next-generation transceiver design, in this paper, we propose a novel framework named \textit{channel deduction} for high-dimensional channel acquisition in multiple-input multiple-output (MIMO)-orthogonal frequency division multiplexing (OFDM) systems. Specifically, it makes use of the outdated channel information of past time slots, performs coarse estimation for the current channel with a relatively small number of pilots, and then fuses these two information to obtain a complete representation of the present channel. The rationale is to align the current channel representation to both the latent channel features within the past samples and the coarse estimate of current channel at the pilots, which, in a sense, behaves as a complementary combination of estimation and prediction and thus reduces the overall overhead. To fully exploit the highly nonlinear correlations in time, space, and frequency domains, we resort to learning-based implementation approaches. By using the highly efficient complex-domain multilayer perceptron (MLP)-mixer for across-space-frequency-domain representation and the recurrence-based or attention-based mechanisms for the past-present interaction, we respectively design two different channel deduction neural networks (CDNets). We provide a general procedure of data collection, training, and deployment to standardize the application of CDNets. Comprehensive experimental evaluations in accuracy, robustness, and efficiency demonstrate the superiority of the proposed approach, which reduces the pilot overhead by up to 88.9\% compared to state-of-the-art estimation approaches and enables continuous operating even under unknown user movement and error propagation.

\end{abstract}

\IEEEpeerreviewmaketitle

\section{Introduction}\label{introduction}
\subsection{Background}
In wireless systems, real-time and accurate channel state information (CSI) is critical for formulating appropriate transceiver patterns and system parameters. The wireless channel varies due to user movement, changes in scatterer distribution, etc., so the wireless system requires channel acquisition in terms of time slots, and the main method used is to estimate the channel by allocating pilots in the resource block of each time slot \cite{pilot}. However, in the sixth-generation (6G) wireless communication systems, along with the use of multiple-input multiple-output (MIMO) technology \cite{MIMO_survey1,MIMO_survey2}, wide bandwidth \cite{mmwave_survey1,mmwave_survey2} and the support for high-frequency communication and high mobility \cite{vision_6G}, the traditional channel acquisition approaches that rely primarily on pilot-based estimation are facing significant challenges. 

On the one hand, multiple antennas and wide bandwidth dramatically increase the channel size, which makes estimation approaches need to allocate more pilots in the resource block, reducing the spectral efficiency. On the other hand, high frequency and high mobility shorten the coherence time, making the acquired channel expire in a shorter time and reducing the period and available time resource of a stable time slot. Therefore, how to timely acquire high-dimensional channels with low signaling overhead has become a critical topic in wireless systems \cite{bigAI6G}.

Although MIMO and orthogonal frequency division multiplexing (OFDM) techniques increase the channel size, the limited distance between antennas and the limited bandwidth between subcarriers result in significant path similarity between subchannels in both space and frequency domains. In addition, path similarity also exists in the channels of neighboring time slots due to the limited movement of users and limited scenario changes in the short period. These similarities provide channels with significant intrinsic correlations in the time-space-frequency domain. Utilizing these correlations becomes the feasibility guarantee and technical key to reducing the overhead of high-dimension channel acquisition.

\begin{figure*}[t]
\centering
  \begin{minipage}[t]{0.46\textwidth}
    \includegraphics[width=\textwidth]{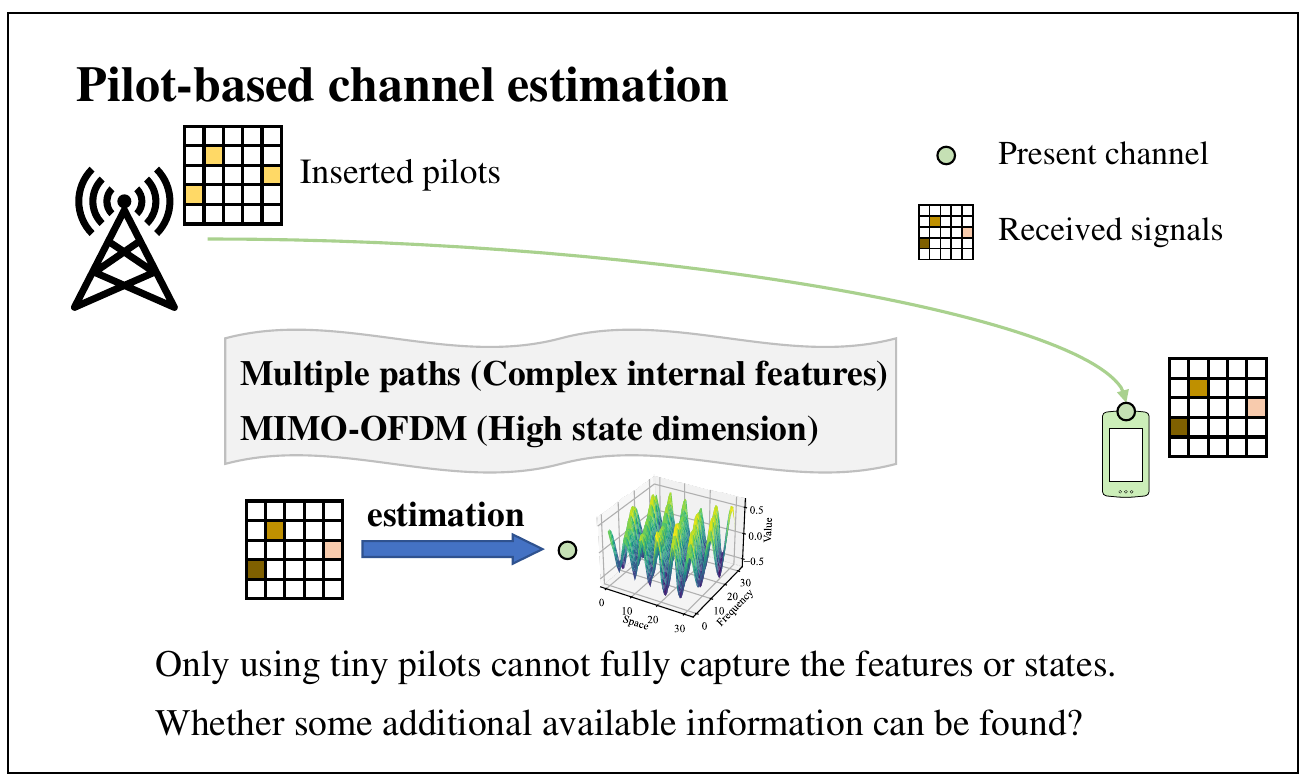}
    \caption*{\footnotesize (a) Illustration of the general channel estimation.}
  \end{minipage}
  \begin{minipage}[t]{0.025\textwidth}
    ~
  \end{minipage}
\begin{minipage}[t]{0.46\textwidth}
    \includegraphics[width=\textwidth]{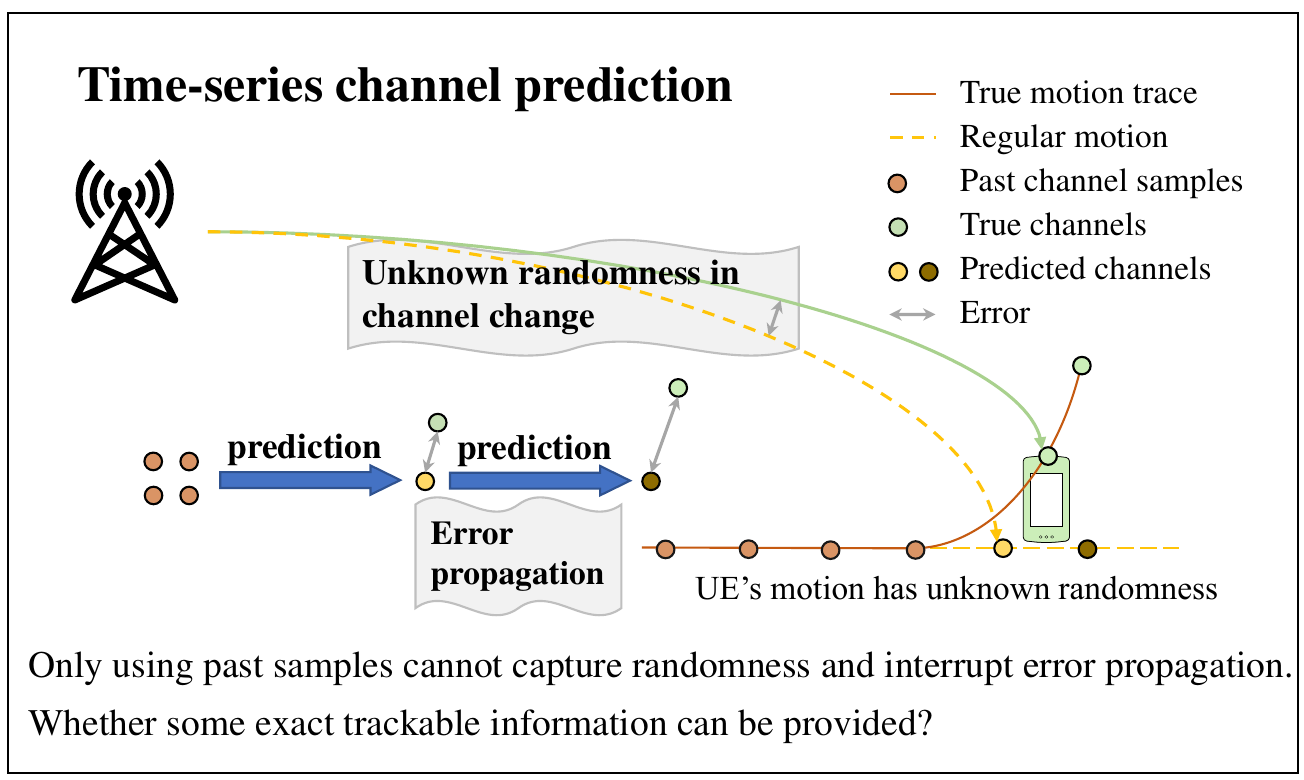}
    \caption*{\footnotesize (b) Illustration of the general channel prediction.}
\end{minipage}

\vspace{1em}

\centering
  \begin{minipage}[t]{0.46\textwidth}
    \includegraphics[width=\textwidth]{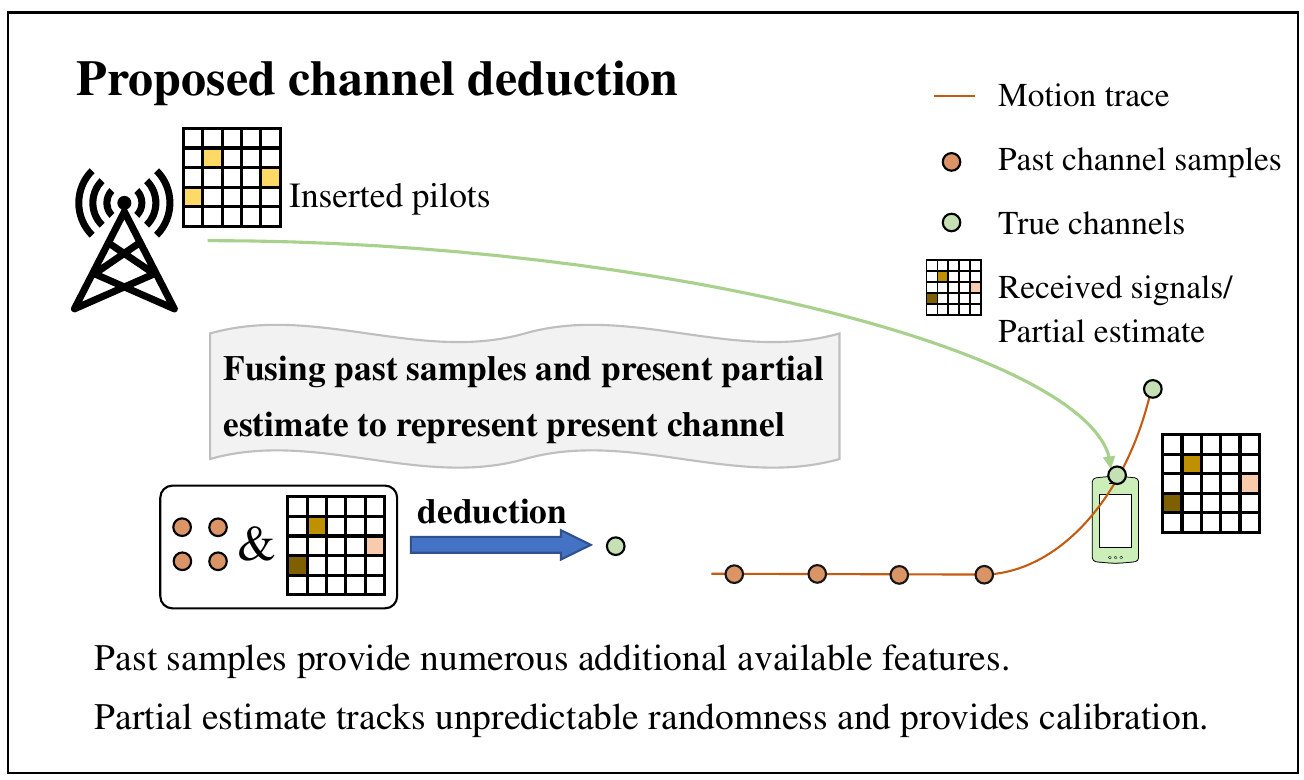}
    \caption*{\footnotesize (c) Proposed channel deduction to address limitations of estimation and prediction.}
  \end{minipage}
  \begin{minipage}[t]{0.025\textwidth}
    ~
  \end{minipage}
\begin{minipage}[t]{0.46\textwidth}
    \includegraphics[width=\textwidth]{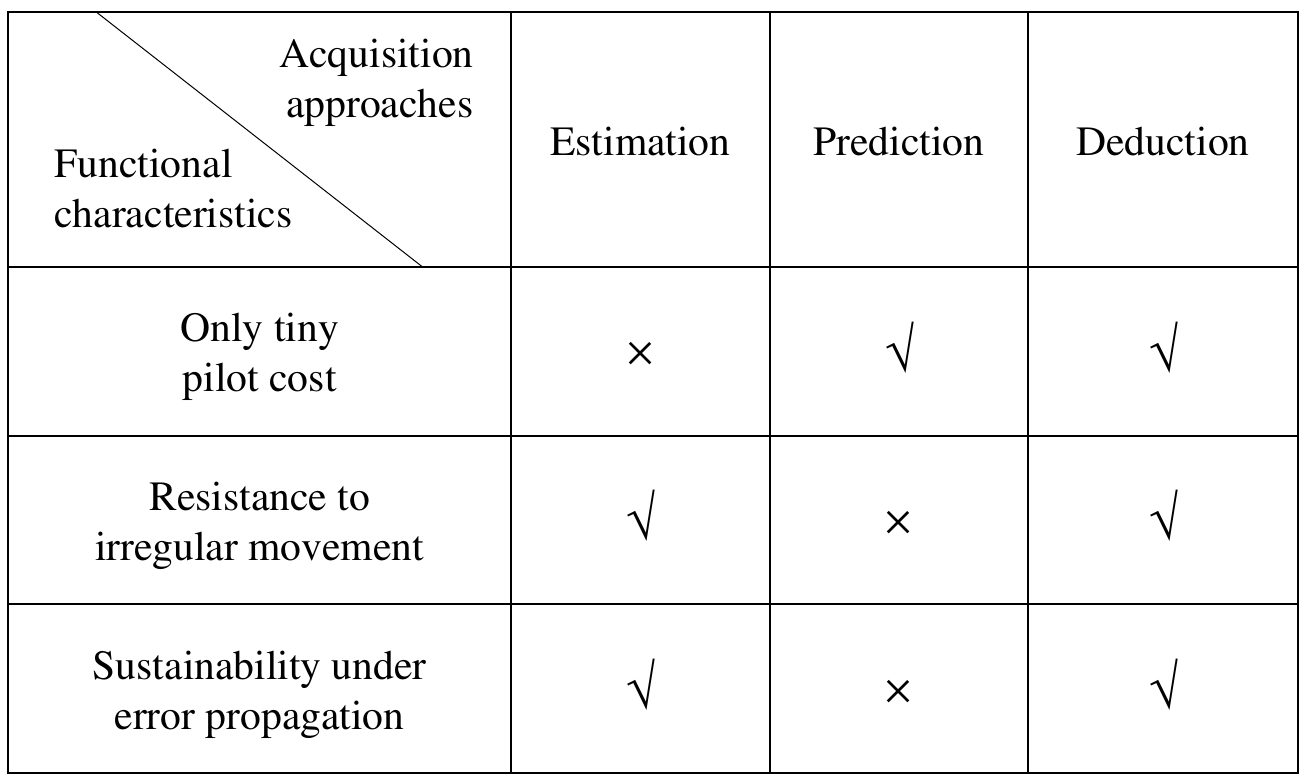}
    \caption*{\footnotesize (d) Functional comparisons between channel estimation, prediction, and deduction.}
\end{minipage}

\caption{\small The motivations and overall architecture of the proposed channel deduction approach.}
\vspace{-0.5em}
\label{fig_introduction}
\end{figure*}

\subsection{Related Works} \label{section1.2}
Some related works have been carried out for this task and can be generally categorized into the following two types: channel estimation enhanced by channel mapping and channel prediction. 

Channel estimation enhanced by channel mapping is that pilots are only used to estimate a small number of present channel features, and the whole channel is mapped from these estimated features based on the internal correlation within the channel. Specifically, in \cite{OFDM_estimation1,OFDM_estimation2,OFDM_estimation3}, the authors used frequency-domain interpolation to obtain all OFDM channels based on the estimated channels of partial subcarriers, constructing pilot-based OFDM channel estimation approaches. Then, with the development of artificial intelligence (AI), deep learning (DL) technology, which is powerful in nonlinear representation, has also been applied to OFDM channel estimation. In \cite{CNN_estimation} and \cite{Res_CNN_estimation}, the authors used convolutional neural networks (CNN) to replace traditional signal processing algorithms for frequency-domain interpolation to improve performance. Moreover, the authors in \cite{CNN_estimation_time} and \cite{time_estimation} respectively introduced a memory cache and a bidirectional long short-term memory (LSTM) \cite{LSTM} to exploit the time correlation within a subgroup/subframe, enhancing the quality and noise resistance of the frequency-domain interpolation. Further, to address the additional space dimension brought by MIMO, the work in \cite{channel_mapping} extended the channel mapping to the space domain and proposed to map the whole MIMO-OFDM channel from subchannels at a subset of antennas and subcarriers. In \cite{cmixer}, the authors proposed a physics-inspired complex-domain MLP-Mixer (CMixer) that improves the multi-layer perceptron (MLP) network in \cite{channel_mapping}, achieving state-of-the-art channel mapping performance. Moreover, \cite{learning_pilot} further proposed pruning the fully-connected layers to reduce the pilot cost based on the MLP networks.

The channel prediction approach differs from estimation in that it does not rely on the signaling resources of the present time slot and consequently has two frameworks: position-based prediction and past channel-based prediction. In \cite{xiao_siren} and \cite{xiao_ode}, the authors used the real-time user position as input to successfully predict channel with high accuracy. However, position-based prediction requires positioning accuracy within a wavelength (millimeter level), which is far from achievable in current systems, especially in practical outdoor scenarios, and thus, such methods can still not be widely applied. In contrast, the mainstream prediction methods are to infer channels of present time slot or even future time slots based on channels of past time slots as inputs. Since past channels are naturally acquired in period of communication with multiple time slots, such methods have strong application potential in cost and framework. In \cite{linear_pred}, the authors autoregressively obtained present CSI from past CSI with linear weighting, and \cite{PCA_pred} improved this autoregression by introducing principal component analysis. However, since time correlation is complicated, the results of such traditional signal processing-based methods are often unsatisfactory. While DL techniques, with their excellent implicit correlation exploitation capability, significantly improve the performance of time-series channel prediction. In \cite{lstm_pred1,lstm_pred2,lstmae_pred}, the authors used recurrent neural networks (RNN), as well as their typical variants, LSTM and gate recurrent unit (GRU) networks, for this time-series task. Besides, the work in \cite{seq2seq_pred} used the sequence to sequence (Seq2Seq) structure to improve the inference for channels of multiple future time slots. In \cite{rnnode_pred}, RNN-ordinary differential equation (ODE) models are used to improve the flexibility of required time slot intervals between past channel samples. Moreover, in \cite{transformer_pred}, the authors applied a transformer model to improve the parallelization of predictions.

\subsection{Motivations and Contributions} \label{section1.3}
{\color{black} Despite the contributions in reducing overhead, both types of the above methods still have limitations, as shown in Fig. \ref{fig_introduction}.} On one hand, the correlations and inputs utilized by most estimation methods are limited to the present time slot only. Then, if the features of the present channel, e.g., multipath components, are complex, feature estimation based on tiny pilots cannot map the present channel with high quality. Although the works in \cite{time_estimation} and \cite{CNN_estimation_time} improved the estimation quality with additional time slots, the utilization of temporal correlation is still limited to time slots within a fixed subframe/subgroup. When beginning new subframes and subgroups, information from previous time slots is still dropped, preventing the continuous utilization of time correlation across the whole time domain, like time-series prediction. These methods are essentially still the estimation but within a larger time block, so while the overhead of some time slots is reduced, the overhead of the whole subframe is still high. On the other hand, although time-series channel prediction saves the pilots of present time slot, due to the movement randomness and the unawareness of scenario changes, some features of present channel cannot be inferred from past information, and the absence of these features makes the prediction often suffer from severe performance degradation. Meanwhile, the autoregressive inference on the time domain inevitably propagates the acquisition errors in past time slots. The current prediction methods lack mechanisms to introduce calibration information, leading to the iterative amplification of the errors.

To address the above limitations, this paper proposes a unified framework with considering both channel mapping and prediction. In fact, these limitations essentially stem from constraints on the sources of information utilized. Based only on the information of present slot or subframe results in a huge requirement for pilots, and based only on the past information prevents unknown feature acquisitions and error calibrations. An approach that can continuously utilize past information to reduce overhead as in prediction, track irregular channel changes, and escape from error propagation as in estimation would complementarily gather the strengths of estimation and prediction to overcome the above limitations. These analyses motivate us to propose a new channel acquisition framework named channel deduction, which autoregressively deduces the referential information from past channels, partially estimates the present channel to obtain unknown features, and then fuses these two information to precisely represent present CSI. Through the DL-based implementation, the proposed channel deduction approach unprecedentedly achieves three functions simultaneously: tiny pilot cost, resistance to movement randomness, and sustainability under error propagation. The main contributions of this paper are summarized as follows:
\begin{itemize}
    \item By analyzing the working mechanisms of estimation and prediction, we reveal the complementarity of the two approaches and thus propose the channel deduction approach that fuses past channels and present partial estimate to represent the present whole channel.
    \item We provide a general process for DL-based channel deduction implementation and further propose two specific channel deduction network (CDNet) schemes with differentiated designs: recurrence-based CDNet (RCDNet) and attention-based CDNet (ACDNet).
    \item We analyze the functional characteristics and complexity of CDNets and provide a general procedure from data collecting, data augmentation, and training to deployment for this new wireless AI technique.
    \item Through extensive simulation experiments in multiple use cases, we comprehensively evaluate the channel acquisition accuracy, robustness, working principle, and application value, illustrating the effectiveness and superiority of the proposed schemes.
\end{itemize}

The remainder of this paper is organized as follows. In Section \ref{system model}, we introduce the system model, including the channel model and channel acquisition framework. Then, our proposed channel deduction approach and related analysis are presented in Section \ref{section3}. Next, Section \ref{section4} shows the performance evaluation of our proposed schemes from various aspects. Finally, Section \ref{conclusion} draws the conclusion.

\section{System Model}\label{system model}
In this section, we introduce the channel model of a MIMO-OFDM system and the framework of the continuous channel acquisition process.

\subsection{Channel Model}\label{section2.1}
{\color{black}We consider a massive MIMO system, where a base station (BS) with $N_\mathrm{t} \gg 1$ antennas serves multiple single-antenna users adopting OFDM modulation with $N_\mathrm{c}$  subcarriers. The channel between the BS and one user is assumed to be composed of $P$ paths, which can be expressed as
\begin{equation}
   {\bf{h}}\left( f \right) = \sum\limits_{p = 1}^P {{\alpha _p}{e^{ - j2\pi f{\tau _p}}}{\bf{a}}(\vec p )},
\label{channel}
\end{equation}
where $f$ is the carrier frequency, ${\alpha _p}$ is the amplitude attenuation, ${\tau _p}$ is the propagation delay, and $\vec p$ is the three-dimensional unit vector of departure direction of the $p$-th path. Furthermore, $\mathbf{a}(\vec p )$ is the array defined as
\begin{equation}
   {\mathbf{a}(\vec p )} = {\left[ {1,{e^{ - j2\pi f {{\vec d_1}}  \cdot \vec p /c}}, \ldots ,{e^{ - j2\pi f {{\vec d_{{N_{\rm{t}}} - 1}}}  \cdot \vec p /c}}} \right]^{\rm{T}}},
\label{array_vector}
\end{equation}
where $\left[ {\vec 0 , {{\vec d_1}} , \ldots , {{\vec d_{{N_{\rm{t}}} - 1}}} } \right]$ is the space vector array, ${{ \vec d_i}}~(i=1,2,\ldots, {{N_{\rm{t}}} - 1})$ represents the  three-dimensional space vector between the $i$-antenna and the first antenna, and $c$ is the speed of light. The transmission distance shift between the $i$-th antenna and the first antenna on $p$-th path can be written as $ {{\vec d_i}}  \cdot \vec p $. Moreover, the overall channel matrix $\mathbf{H} \in {\mathbb{C}^{{N_\mathrm{t}}{\rm{ \times }}{N_\mathrm{c}}}}$ between the BS and this user can be expressed as, 
\begin{equation}
   \mathbf{H} = [\mathbf{h}({f_1}),\mathbf{h}({f_2}), \ldots ,\mathbf{h}({f_{{N_\mathrm{c}}}})],
\label{channel matrix}
\end{equation}
where $f_1$, $f_2$, $\cdots$, ${f_{{N_\mathrm{c}}}}$ denote the subcarrier frequencies of all ${N_\mathrm{c}}$ subcarriers. }

\subsection{Channel Acquisition}\label{section2.2}
\begin{figure}[htbp]
  \centering
  \includegraphics[width=0.96\linewidth]{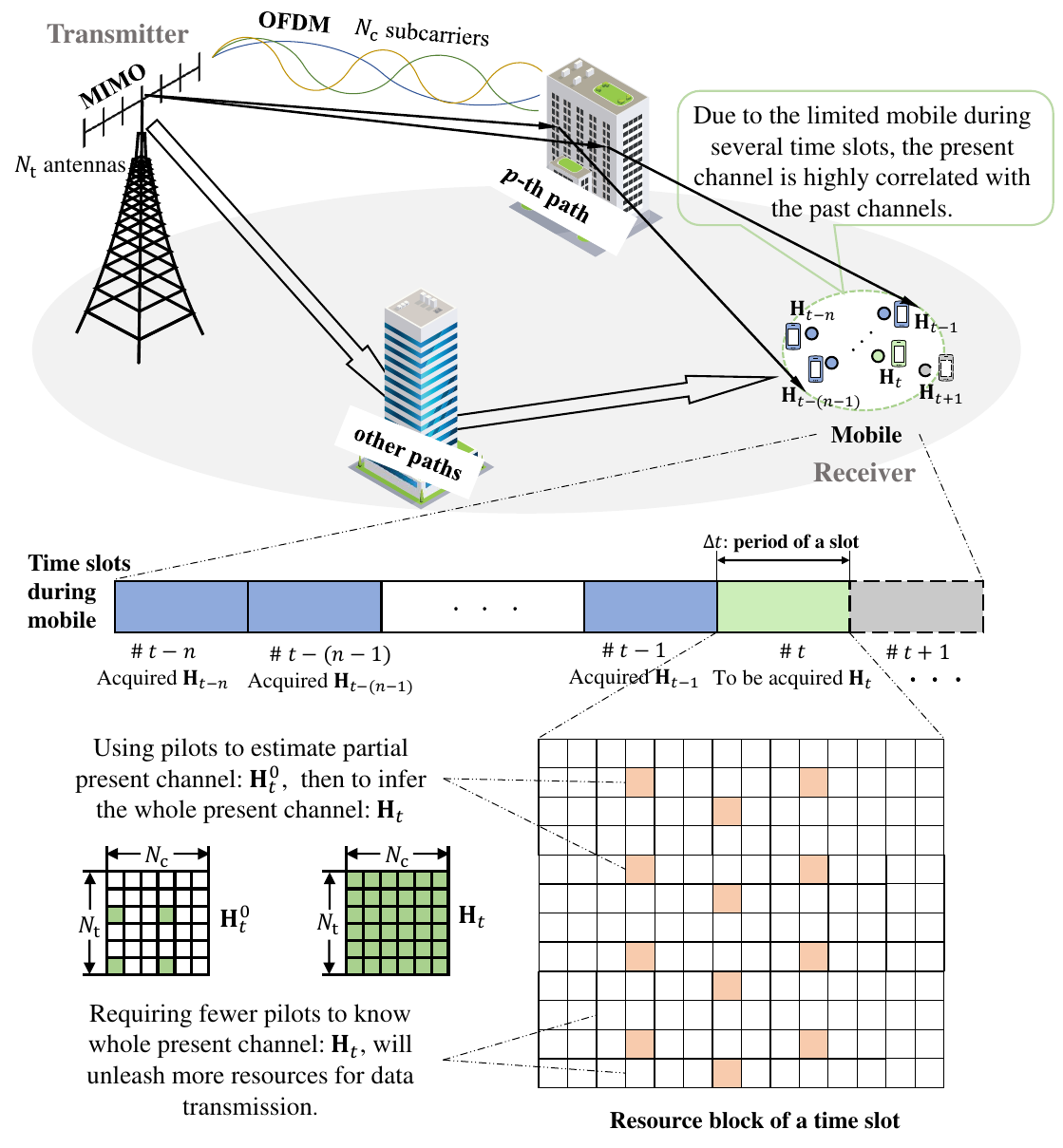}
  \caption{\small \color{black} Mobile channel in a MIMO-OFDM system.}
  \vspace{-1em}
  \label{channel_model_fig}
\end{figure}
Accurately obtaining the current channel state is crucial for establishing high-gain transceiver schemes, signal detection, and numerous other wireless tasks. During the communication process, the channel state changes continuously as the user moves. Within a coherence time, the channel state can be assumed to be approximately steady, whereas over a coherence time, the channel state becomes expired and unreliable. Therefore, the current wireless system establishes a mechanism to continuously acquire channels regarding time slots, as shown in Fig. \ref{channel_model_fig}. Specifically, inserting several pilots into the resource block of a time slot and estimating the current CSI based on the received signal, which can be written as,
\begin{equation}
    {\bf{y}} = {\bf{H}}{\bf{x}} + {\bf{n}},
\end{equation}
where $\bf{y}$ is the received signal, $\bf{x}$ is the transmitted signal, both known by transmitter and receiver, and ${\bf{n}}$ is the noise. Then, the estimated channel can be used for the transmission in this time slot. 

However, due to the anticipated support for massive MIMO, wide bandwidth, millimeter wave (mmWave), and high-speed mobility in next-generation wireless networks, existing channel acquisition methods that rely purely on pilots face significant challenges.
First, massive MIMO and wide bandwidth significantly increase the channel size, which leads to a rise in pilot overhead for estimating the whole MIMO-OFDM channel. Additionally, since coherence time is inversely proportional to speed and frequency, high mobility and high frequency accelerate channel expiration, reducing the resources available for channel acquisition. Therefore, how to acquire mobile MIMO-OFDM channels at a low cost has become a crucial problem in next-generation wireless networks.

\section{Proposed Channel Deduction Framework}\label{section3}

\subsection{Motivations and Framework of Channel Deduction} \label{section3.1}

\subsubsection{Existing Channel Acquisition Approaches and Limitations}
Two typical approaches have been proposed for low-cost mobile MIMO-OFDM channel acquisition. One approach is channel estimation enhanced by channel mapping in the space-frequency domain \cite{channel_mapping,cmixer}. This approach first estimates present state information of partial antennas and subcarriers, then maps it to the whole MIMO-OFDM CSI, which can be written as follows,
{\color{black}
\begin{align}
{{\bf{H}}_t} &= {g_{{\rm{cm}}}}\left( {{\bf{H}}_t^0} \right), 
\end{align}
where ${{\bf{H}}_t}$ is the present channel to be acquired, ${g_{{\rm{cm}}}}:{{\mathbb{C}}^{N_{\rm{t}}^0{\rm{ \times }}N_{\rm{c}}^0}} \to {{\mathbb{C}}^{{N_{\rm{t}}}{\rm{ \times }}{N_{\rm{c}}}}}$ is the channel mapping function, }${{\bf{H}}_t^0}$ is the known CSI of several antennas and subcarriers estimated through pilots. And ${{\bf{H}}_t^0}$ is also written as ${{\bf{H}}_t}[\Omega ]$, where $\Omega$ is the selected small subset in space and frequency. 
This approach leverages the space-frequency correlation in MIMO-OFDM CSI, reducing the channel size to be estimated from ${N_{\rm{t}}}{\rm{ \times }}{N_{\rm{c}}}$ to $\left| \Omega  \right|$.

However, enhancements based on channel mapping do not essentially address the limitations of channel estimation since all known inputs used for present channel acquisition still require pilots. {\color{black} When the multipath features of present channel are complex and dynamic, a large size ${{\bf{H}}_t^0}$ is still needed to capture enough features for high-quality representation of ${{\bf{H}}_t}$. Therefore, only relying on estimate is generally inefficient and inadequate.}

Different from relying on the information of present time slots, another approach, channel prediction, acquires present CSI based on CSI of past time slots. This approach is motivated by the time correlation between past and present channels. During several time slots, the user's motion is limited.
This spatial proximity leads the channels of neighboring time slots to be highly similar in terms of large-scale features such as multipath structure, AoD of each path, and attenuation of each path, arising the time correlation. Further, channel prediction is to realize the following function based on time correlation,
{\color{black}
\begin{equation}
    {{\bf{H}}_t} = {{g}_{{\rm{cp}}}}\left( {{{\bf{H}}_{t - n}}, \ldots ,{{\bf{H}}_{t - 1}}} \right),
\end{equation}
where ${g_{{\rm{cp}}}}:(\underbrace {{{\mathbb{C}}^{{N_{\rm{t}}}{\rm{ \times }}{N_{\rm{c}}}}}, \ldots ,{{\mathbb{C}}^{{N_{\rm{t}}}{\rm{ \times }}{N_{\rm{c}}}}}}_{n~{\rm{ terms}}}) \to {{\mathbb{C}}^{{N_{\rm{t}}}{\rm{ \times }}{N_{\rm{c}}}}}$ is channel prediction function}, ${{\bf{H}}_{t - i}}~(i=1,2,\ldots,n)$ is the CSI on ${(t - i)}$-th time slot. Further, since the CSI of past time slots ${{{\bf{H}}_{t - n}}, \ldots ,{{\bf{H}}_{t - 1}}}$ is naturally already acquired when the communication process proceeds to $t$-th time slot, ${{\bf{H}}_t}$ can be predicted based on ${{g}_{{\rm{cp}}}}$ without pilot overhead. Besides, this channel acquisition process can be continued autoregressively, i.e., ${{\bf{H}}_{t{\rm{ + }}k}} = {{g}_{{\rm{cp}}}}\left( {{{\bf{H}}_{t +k- n}}, \ldots ,{{\bf{H}}_{t+k-1}}} \right),~k=1,2,\ldots$, benefiting from the persistent time correlation.

{\color{black} However, the past and the present do not always have a strict causal relationship. For example, where the user will move in the next time slot is not fully predictable only based on past positions. In other words, past and present spatial positions sometimes have only a proximity rather than a deterministic relationship. Given that user position is closely related to user channel, such uncertainty specifically results in a wide range of possibilities for fast-change small-scale features such as phase of present channel, which makes ${{\bf{H}}_t}$ cannot be fully obtained only based on ${{{\bf{H}}_{t - n}}, \ldots ,{{\bf{H}}_{t - 1}}}$.}
Then, ${{g}_{{\rm{cp}}}(\cdot)}$ is thus often not a learnable function, resulting in large errors when meeting irregular motion. Meanwhile, the acquired error also accumulates in the autoregressive predictions, deteriorating performance continuously. For these reasons, the channel prediction approach cannot be solely relied on in practical systems despite its great potential to save pilots. {\color{black} Instead, real-time (or on-the-fly) channel measurements are necessary to further explore the present channel characteristics.}

\subsubsection{Overview of Channel Deduction}
\begin{figure}[htbp]
  \centering
  \includegraphics[width=0.98\linewidth]{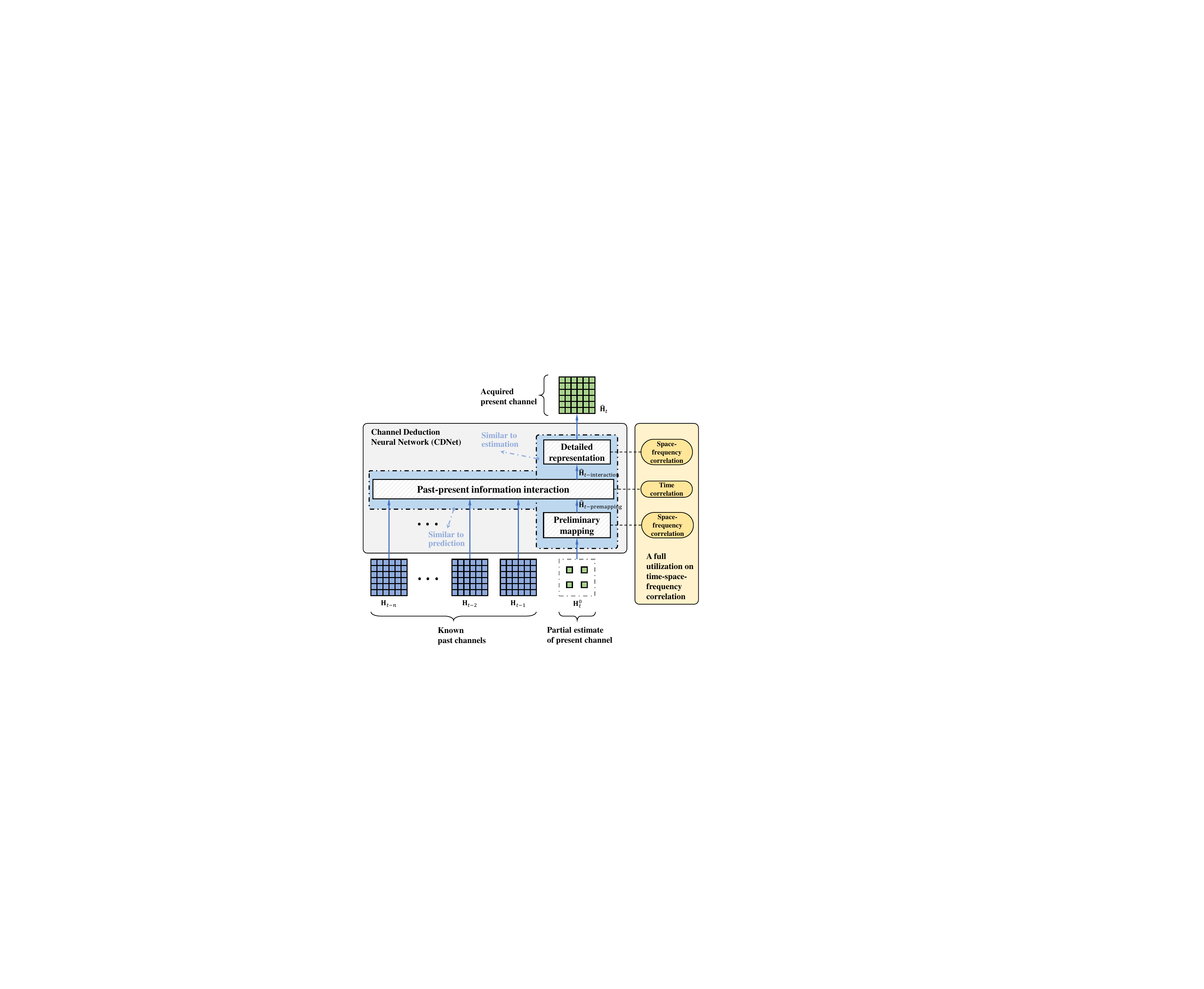}
  \captionsetup{labelfont={color=black}, textfont={color=black}}
  \caption{\small \color{black}The overall structure of CDNet.}
  \vspace{-0.3em}
  \label{CDNet_fig}
\end{figure}

While the limitations of prediction and estimation motivate us to explore more efficient methods for channel acquisition, exploiting time and space-frequency correlation to acquire channel cost-effectively is still very instructive. In fact, as it is possible to approximately infer the large-scale features of ${{\bf{H}}_t}$ from ${{{\bf{H}}_{t - n}}, \ldots ,{{\bf{H}}_{t - 1}}}$ and obtain other unknown features, including small-scale features, from ${{\bf{H}}_t^0}$, if an approach can effectively fuse these two information to represent the present channel, it will be a desirable approach for channel acquisition. Specifically, this approach can be written as follows,
{\color{black} 
\begin{equation}
{{\bf{H}}_t} = {{g}_{{\rm{cd}}}}\left( {{{\bf{H}}_{t - n}}, \ldots ,{{\bf{H}}_{t - 1}},{\bf{H}}_t^0} \right),
\label{eq_cd}
\end{equation}
where ${g_{{\rm{cd}}}}:(\underbrace {{{\mathbb{C}}^{{N_{\rm{t}}}{\rm{ \times }}{N_{\rm{c}}}}}, \ldots ,{{\mathbb{C}}^{{N_{\rm{t}}}{\rm{ \times }}{N_{\rm{c}}}}}}_{n~{\rm{ terms}}},{{\mathbb{C}}^{N_{\rm{t}}^0{\rm{ \times }}N_{\rm{c}}^0}}) \to {{\mathbb{C}}^{{N_{\rm{t}}}{\rm{ \times }}{N_{\rm{c}}}}}$. }Since this approach deduces general features from channel samples of neighboring time slots and aligns these features to the present channel representation with the help of partial estimate on the present, we name it ``\textit{channel deduction}''.
The $\rm{g}_{\rm{cd}}(\cdot)$ in Eq. \eqref{eq_cd} is the channel deduction function. And beautifully, ${{\bf{H}}_{t - n}}, \ldots ,{{\bf{H}}_{t - 1}}$ and ${\bf{H}}_t^0$ will complement each other to address the limitations of estimation and prediction. On one hand, compared to channel estimation, channel deduction additionally utilizes the referable information from ${{\bf{H}}_{t - n}}, \ldots ,{{\bf{H}}_{t -1}}$, which is especially valuable when the size of ${\bf{H}}_t^0$ is not large. It will bring a higher-quality channel acquisition under the same pilot overhead. On the other hand, by introducing ${\bf{H}}_t^0$, the exact estimation of present time slot, channel deduction can capture the movement randomness which cannot be predicted only based on ${{\bf{H}}_{t - n}}, \ldots ,{{\bf{H}}_{t - 1}}$. Meanwhile, ${\bf{H}}_t^0$, which comes from an exact estimation rather than time-correlation-based inference, can provide continuous calibration information during the autoregressive process to restrain the error propagation.

\subsection{\color{black} DL-Enabled Channel Deduction Implementation} \label{section3.2}
To obtain the function ${{g}_{{\rm{cd}}}(\cdot)}$ in Eq. \eqref{eq_cd}, it is necessary to mine and utilize the time correlation between ${{\bf{H}}_{t - n}}, \ldots ,{{\bf{H}}_{t - 1}}$ and ${\bf{H}}_t$ as well as the space-frequency correlation between ${\bf{H}}_t^0$ and ${\bf{H}}_t$. Since CSI is a complex coupling of multipath channel responses, these time-space-frequency correlations, which originate from single-path similarities, are often implicit. Mining such correlations is often difficult using traditional signal processing methods, while DL methods with excellent implicit feature mining and data representation capabilities are outstandingly competitive. Therefore, we use neural networks to realize channel deduction.

\begin{figure}[htbp]\centering
	\subfigure[A layer of CMixer]{
		\includegraphics[height=0.4\textwidth]{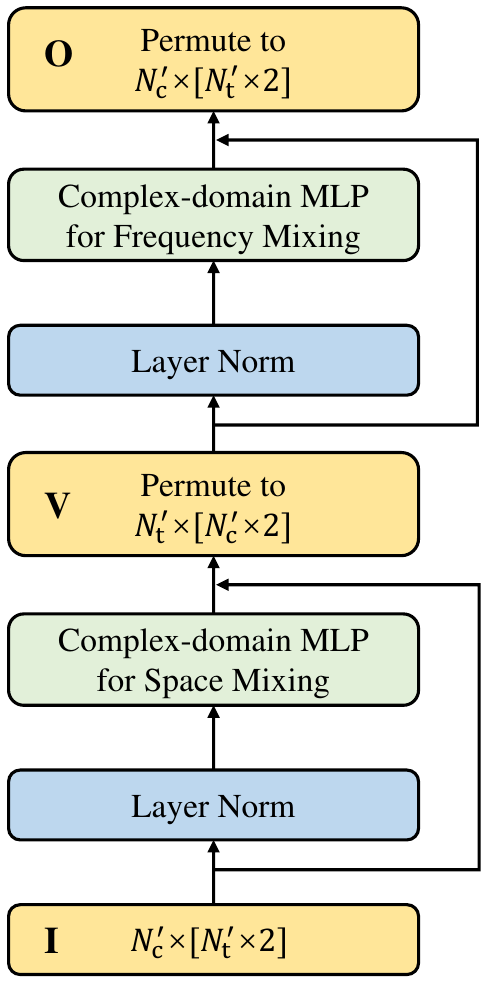}
	}
	\subfigure[A CMixer module made of $K$ stacked layers] {
		\includegraphics[height=0.4\textwidth]{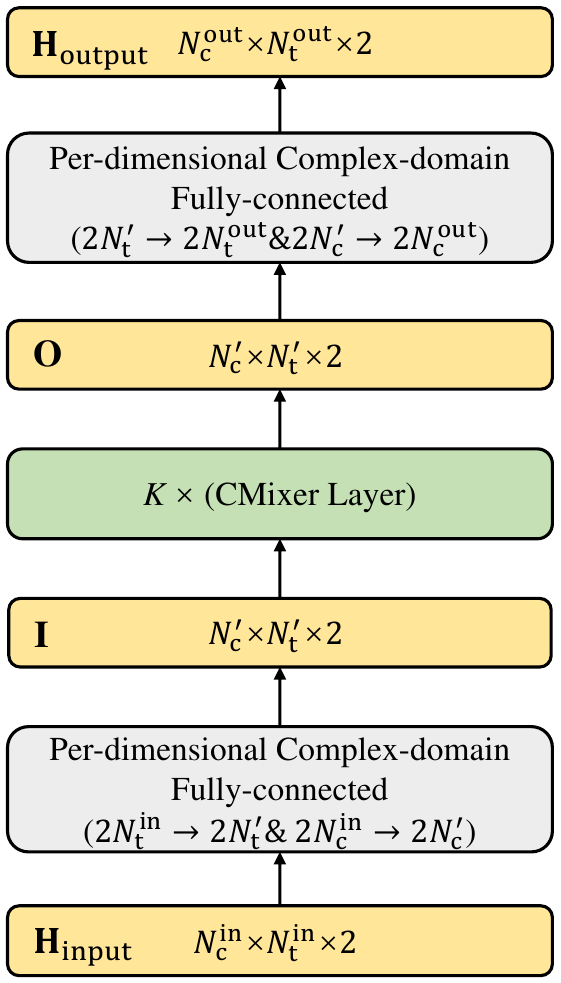}
	}
	\caption{\small The specific network structure of CMixer \cite{cmixer}. The CMixer is used to implement the ``Preliminary mapping'' and ``Detailed representation'' module in CDNet. }
	\vspace{-0.3em}
\label{cmixer_fig}
\end{figure}

\begin{figure*}[htbp]
  \centering
  \includegraphics[width=0.72\linewidth]{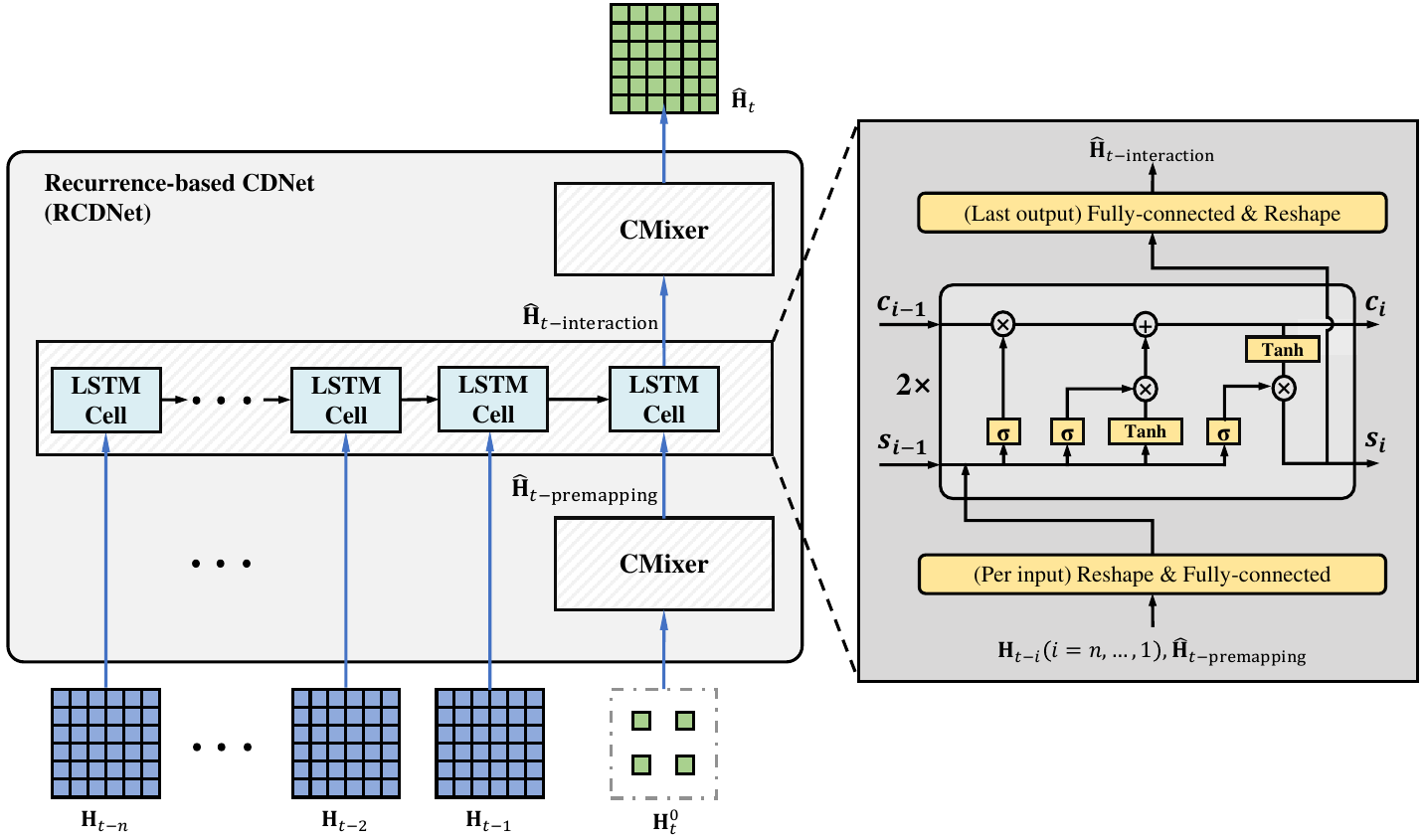}
  \caption{\small The specific network structure of the proposed RCDNet.}
  \vspace{-1em}
  \label{RCDNet_fig}
\end{figure*}

{\color{black}Functionally, channel deduction is a combination of prediction and estimation, which requires both mining the time-domain correlation between ${\bf{H}}_{t}$ and ${{\bf{H}}_{t - n}}, \ldots ,{{\bf{H}}_{t - 1}}$ as in prediction, and exploiting the space-frequency correlation between ${\bf{H}}_{t}^0$ and ${\bf{H}}_{t}$ as in mapping-enhanced estimation, which inspired us to design CDNet in a way that couples a typical time-series prediction network and a space-frequency mapping network. As shown in Fig. \ref{CDNet_fig}, the overall structure of the proposed CDNet includes three modules: preliminary mapping, past-present information interaction, and detailed representation. In a sense, `past-present information interaction' can be analogized to the mining of time-domain correlations in prediction. Meanwhile, `preliminary mapping' and `detailed representation' are similar to mining space-frequency correlations in mapping. By coupling these three modules in the manner shown in Fig. \ref{CDNet_fig} and then co-training them end-to-end, one can drive CDNet to realize the functional combination of prediction and estimation. Moreover, it can be intuitively obtained that the computational and parameter complexity of CDNet is approximately the sum of the prediction network and the estimation network and, thus, is in the same order of magnitude as the larger of the two. The overall computation and workflow of CDNet is as follows.
$N_{\rm{t}}^0{\rm{ \times }}N_{\rm{c}}^0 \times 2$-size ${\bf{H}}_t^0$ is first mapped to ${N_{\rm{t}}}{\rm{ \times }}{N_{\rm{c}}} \times 2$-size ${{\bf{H}}_{t - {\rm{premapping}}}}$ based on the space-frequency correlation ($\times 2$ denotes transfer complex-value to real-value), since the same dimension as ${{\bf{H}}_{t - n}}, \ldots ,{{\bf{H}}_{t - 1}}$ facilitates the following past-present information interaction. Then, CDNet supplements ${{\bf{H}}_{t - {\rm{premapping}}}}$ with usable features from ${{\bf{H}}_{t - n}}, \ldots ,{{\bf{H}}_{t - 1}}$ to obtain a more fully informative ${{\bf{H}}_{t - {\rm{interaction}}}}$. Finally, CDNet represents the targeted ${\widehat {\bf{H}}_t}$ based on ${{\bf{H}}_{t - {\rm{interaction}}}}$  according to the inherent space-frequency characteristics of the MIMO-OFDM channel. }

Next, we introduce the specific network structure used in this paper for the implementation of CDNet. {\color{black} Firstly, for space-frequency correlation-based representation, including the ``Preliminary mapping'' and ``Detailed representation'' modules in Fig. \ref{CDNet_fig}, the work in \cite{cmixer} has provided an advanced CMixer scheme. The brief network structure of CMixer is shown in Fig. \ref{cmixer_fig}. Through the unique interleaved space and frequency learning, the CMixer can learn to map a $N_{\rm{c}}^{{\rm{in}}}{\rm{ \times }}N_{\rm{t}}^{{\rm{in}}} \times 2$-size channel matrix to a $N_{\rm{c}}^{{\rm{out}}}{\rm{ \times }}N_{\rm{t}}^{{\rm{out}}} \times 2$-size channel matrix with tightening the coupling of space and frequency. In addition, it also has the following three advantages: good structural flexibility to customize the appropriate network structure according to the input and output sizes; lightweight parameter scale and computational complexity; and excellent convergence on channel-related tasks, which facilitates its stacking with other learning modules. Based on the above properties, it is a suitable module for the MIMO-OFDM channel representation task, similar to that the CNN module is suitable for efficient image representation.} The principles and details of CMixer can be found in \cite{cmixer}. In this paper, we use a $K_1$-layer CMixer, whose input dimension is defined as $N_{\rm{t}}^0{\rm{ \times }}N_{\rm{c}}^0 \times 2$ and output dimension is defined as $N_{\rm{t}}{\rm{ \times }}N_{\rm{c}}\times2$, to map ${\bf{H}}_t^0$ to ${{\bf{H}}_{t - {\rm{premapping}}}}$, realizing the ``Preliminary mapping'' module. Moreover, we use a $K_2$-layer CMixer, whose input dimension is defined as $N_{\rm{t}}{\rm{ \times }}N_{\rm{c}} \times 2$ and output dimension is defined as $N_{\rm{t}}{\rm{ \times }}N_{\rm{c}}\times2$, to represent ${\widehat {\bf{H}}_t}$ from ${{\bf{H}}_{t - {\rm{premapping}}}}$, realizing the ``Detailed representation'' module.

Next, we present the specific implementation for the ``Past-present information interaction'' module, mainly used to mine and exploit time correlation. Unlike the space-frequency correlation, which comes from the antenna array form and the subcarrier allocation pattern, the time correlation comes from the user mobility with diverse patterns. Therefore, it is more open and challenging to learn time correlation efficiently. The next subsection provides a sequential move-based and a referable neighborhood-based perspective on time correlation, yielding recurrence-based CDNet and attention-based CDNet, respectively. The two methods are both effective in accomplishing channel deduction while exhibiting differentiated characteristics.

\begin{figure*}[htbp]
  \centering
  \includegraphics[width=0.65\linewidth]{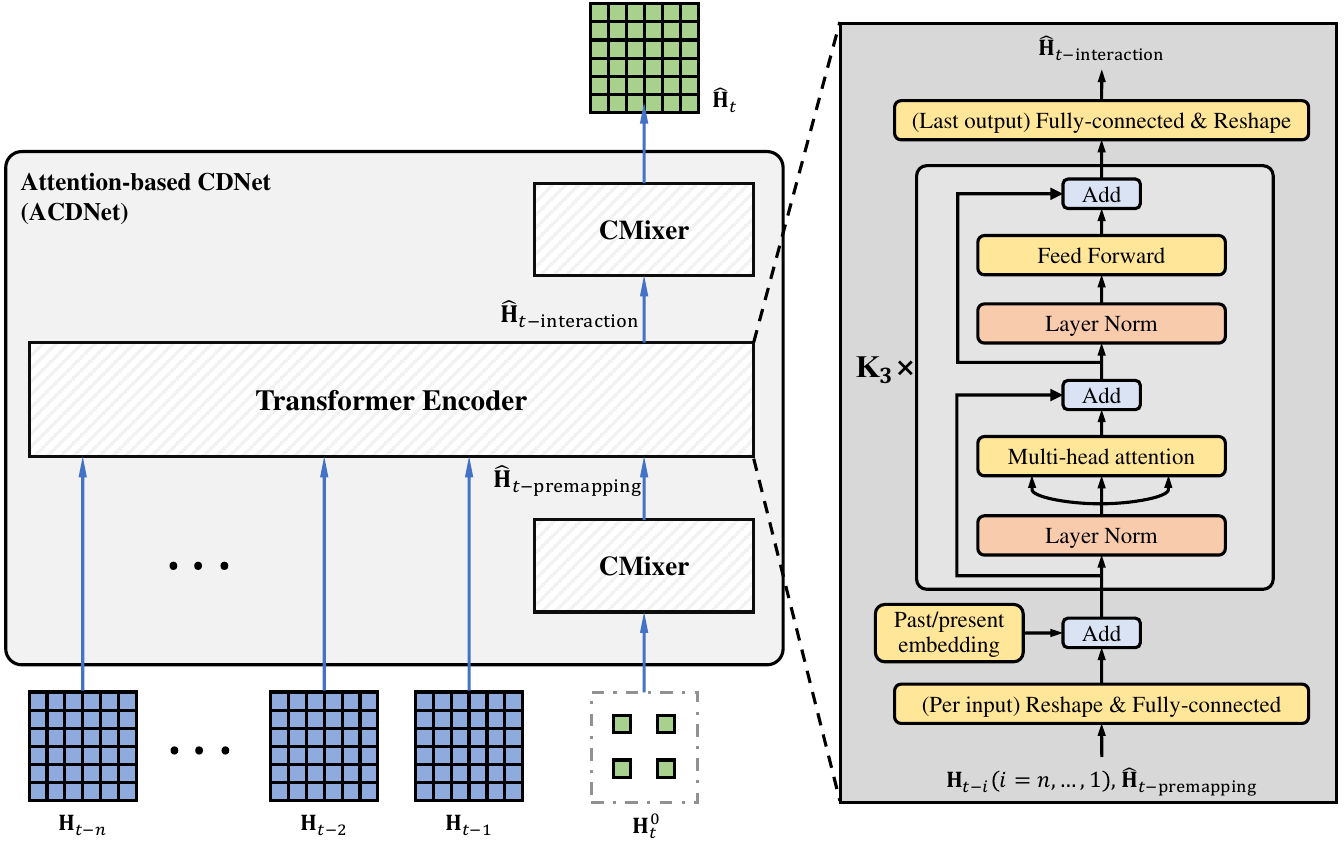}
  \caption{\small The specific network structure of the proposed ACDNet.}
  \vspace{-1em}
  \label{ACDNet_fig}
\end{figure*}

\subsection{RCDNet and ACDNet}\label{section3.3}
\subsubsection{\color{black}Recurrence-Based CDNet} 
Since ${{\bf{H}}_{t - n}}, \ldots ,{{\bf{H}}_{t - 1}}$, ${{\bf{H}}_{t - {\rm{premapping}}}}$ are naturally discrete samplings of a continuously varying physical information series, time correlation can be exploited through sequential interaction. An effective information extraction method for sequence data is the recurrent neural network (RNN) \cite{deep_learning}, which can pass and accumulate effective features to infer the target result by recursion in the sequence direction. Through recurrence-based learning, numerous referable features can be summarized from ${{\bf{H}}_{t - n}}, \ldots ,{{\bf{H}}_{t - 1}}$ and passed to ${{\bf{H}}_{t - {\rm{premapping}}}}$, coupling to generate ${{\bf{H}}_{t - {\rm{interaction}}}}$. Based on this perspective, we use an RNN to realize the past-present information interaction in CDNet, building RCDNet, whose specific structure is shown in Fig. \ref{RCDNet_fig}.

The ``Past-present information interaction'' module of RCDNet includes a $2{N_{\rm{t}}}{N_{\rm{c}}} \to S~(S<2{N_{\rm{t}}}{N_{\rm{c}}})$ fully-connected layer for dimensionality reduction, a 2-layer LSTM \cite{LSTM}, and a $S \to 2{N_{\rm{t}}}{N_{\rm{c}}}$ fully-connected layer for dimensionality recovery. The reason for using fully-connected layers for dimension reduction and recovery is that these $2{N_{\rm{t}}}{N_{\rm{c}}}$-size channel matrices are highly sparse. Thus, past-present information interaction can be accomplished under a relatively low dimension $S$, which results in a low complexity. In addition, LSTM is a typical RNN variant with a gate mechanism and cell memory that further improves the efficiency of information passing between cells. The input and hidden sizes of the used LSTM are both $S$. The calculation process from $\{{{\bf{H}}_{t - n}}, \ldots ,{{\bf{H}}_{t - 1}}, {{\bf{H}}_{t - {\rm{premapping}}}}\}$ to ${{\bf{H}}_{t - {\rm{interaction}}}}$ in RCDNet is as follows. First, each of $\{{{\bf{H}}_{t - n}}, \ldots ,{{\bf{H}}_{t - 1}}, {{\bf{H}}_{t - {\rm{premapping}}}}\}$ is reshaped to a vector, and then is downscaled into an $S$-size vector by the $2{N_{\rm{t}}}{N_{\rm{c}}} \to S$ fully-connected layer. Then, these $n+1$ vectors are sequentially input into the corresponding LSTM cell in time order. Finally, the output of the last cell is recovered to ${N_{\rm{t}}}\times{N_{\rm{c}}}\times2$-size by the $S \to 2{N_{\rm{t}}}{N_{\rm{c}}}$ fully-connected layer and reshape operation to obtain ${{\bf{H}}_{t - {\rm{interaction}}}}$. 
Summarizing Fig. \ref{RCDNet_fig}, the CMixer-based `Preliminary mapping' and `Detailed representation' introduced in Section \ref{section3.2} and the above LSTM-based sequential ``Past-present information interaction'' constitute an efficient implementation of channel deduction, RCDNet.

\subsubsection{\color{black}Attention-Based CDNet}
Not only is a time sequence, $\{{{\bf{H}}_{t - n}}, \ldots ,{{\bf{H}}_{t - 1}}, {{\bf{H}}_{t - {\rm{premapping}}}}\}$ is also a set of highly correlated time neighbors. Therefore, in addition to sequential information accumulation, it is also possible for past and present information to interact with each other by mutual queries and adaptive feature complementation. Specifically, at the standpoint of ${{\bf{H}}_{t - {\rm{premapping}}}}$, other CSI $\{{{\bf{H}}_{t - n}}, \ldots ,{{\bf{H}}_{t - 1}}\}$ are all available neighbor information, and it is possible to query them and draw useful information to yield a more informative ${{\bf{H}}_{t - {\rm{interaction}}}}$. Similarly, ${{\bf{H}}_{t - i}}~(i=1,2,\dots,n)$ can query $\{{{{\bf{H}}_{t - j}}~(j \ne i), {{\bf{H}}_{t - {\rm{premapping}}}}}\}$ to represent valuable features and improve the efficiency of information interaction. Attention mechanism \cite{attention} is an effective means to realize query-based interaction.
And for mutual querying within a sequence, the transformer encoder based on self-attention \cite{transformer} is a typical learning structure leveraging attention connections between sequence data. 
Through attention-based learning, the features from $\{{{\bf{H}}_{t - n}}, \ldots ,{{\bf{H}}_{t - 1}}, {{\bf{H}}_{t - {\rm{premapping}}}}\}$ can be deduced and fused to obtain ${{\bf{H}}_{t - {\rm{interaction}}}}$.  Based on this perspective, we use a transformer encoder to realize the past-present information interaction in CDNet, building ACDNet, whose specific structure is shown in Fig. \ref{ACDNet_fig}.

\begin{table*}[htbp]\footnotesize
\renewcommand{\arraystretch}{1.8}
\caption{\small Comparisons of RCDNet and ACDNet.}
\vspace{-0.8em}
\begin{center}
\begin{tabular}{ c c c c}
\toprule
Schemes & Computation complexity & Degree of past-present interaction & Functional advantages  \\ 
\toprule
RCDNet & $\mathcal{O}(S^2n+{N_{\rm{c}}}{N_{\rm{t}}}^2+{N_{\rm{t}}}{N_{\rm{c}}}^2+{N_{\rm{t}}}{N_{\rm{c}}}Sn)$ & Sequential receptive field (Robust) & Stability under error propagation \\ 
ACDNet & $\mathcal{O}(S^2n+Sn^2+{N_{\rm{c}}}{N_{\rm{t}}}^2+{N_{\rm{t}}}{N_{\rm{c}}}^2+{N_{\rm{t}}}{N_{\rm{c}}}Sn)$& Global receptive field (Sensitive) &  Performance in quasi-static case \\ 
\toprule
\end{tabular}
\end{center}
\vspace{-1.6em}
\label{table_compare}
\end{table*}

The ``Past-present information interaction'' module of ACDNet includes a $2{N_{\rm{t}}}{N_{\rm{c}}} \to S$ fully-connected layer for dimensionality reduction, a $K_3$-layer transformer encoder \cite{transformer}, and a $S \to 2{N_{\rm{t}}}{N_{\rm{c}}}$ fully-connected layer for dimensionality recovery. The structure and motivation of the fully connected layers in ACDNet are the same as in RCDNet. The transformer encoder used in this paper employs the pre-norm variant \cite{pre_norm}, since this change strengthens the residual connectivity, improve the stability and efficiency of  representation learning. The input size and hidden size of the used transformer are both $S$. The calculation process from $\{{{\bf{H}}_{t - n}}, \ldots ,{{\bf{H}}_{t - 1}}, {{\bf{H}}_{t - {\rm{premapping}}}}\}$ to ${{\bf{H}}_{t - {\rm{interaction}}}}$ in ACDNet is as follows. First, each of $\{{{\bf{H}}_{t - n}}, \ldots ,{{\bf{H}}_{t - 1}}, {{\bf{H}}_{t - {\rm{premapping}}}}\}$ is reshaped to a vector, and then is downscaled into an $S$-size vector by the $2{N_{\rm{t}}}{N_{\rm{c}}} \to S$ fully-connected layer. Then, since the attention mechanism is unable to perceive the order information of input, we add different time embeddings for vectors from the past and the present to distinguish them, i.e., adding an $S$-size learnable past embedding to each vector from $\{{{\bf{H}}_{t - n}}, \ldots ,{{\bf{H}}_{t - 1}}\}$ and adding another $S$-size learnable present embedding to the vector from ${{\bf{H}}_{t - {\rm{premapping}}}}$. After the embedding operation, the new $n+1$ vectors are input into the transformer encoder in parallel. Finally, the last vector of the output sequence of the transformer encoder, the output corresponding to ${{\bf{H}}_{t - {\rm{premapping}}}}$,  is recovered to ${N_{\rm{t}}}\times{N_{\rm{c}}}\times2$-size by the $S \to 2{N_{\rm{t}}}{N_{\rm{c}}}$ fully-connected layer and reshape operation to obtain ${{\bf{H}}_{t - {\rm{interaction}}}}$. 
Summarizing Fig. \ref{ACDNet_fig}, the CMixer-based ``Preliminary mapping'' and ``Detailed representation'' introduced in Section \ref{section3.2} and the above transformer encoder-based complementary sequential ``Past-present information interaction'' constitute another efficient implementation of channel deduction, ACDNet.

\subsubsection{Property and Complexity Analysis of RCDNet and ACDNet}

{\color{black} In terms of magnitude order, the computational complexity of RCDNet is $\mathcal{O}(S^2n+{N_{\rm{c}}}{N_{\rm{t}}}^2+{N_{\rm{t}}}{N_{\rm{c}}}^2+{N_{\rm{t}}}{N_{\rm{c}}}Sn)$, which exactly equal to the sum of LSTM-based prediction $\mathcal{O}(S^2n+{N_{\rm{t}}}{N_{\rm{c}}}Sn)$ and CMixer-based estimation $\mathcal{O}({N_{\rm{c}}}{N_{\rm{t}}}^2+{N_{\rm{t}}}{N_{\rm{c}}}^2)$. This confirms the general complexity analysis based on the overall architecture in Section \ref{section3.2}, the computational complexity of CDNet is approximately the sum of the prediction network and the estimation network.} In addition, among the RCDNet and ACDNet, the different learning structures bring differentiated characteristics to them. The first difference lies in the computational complexity. The CMixers and fully-connected layers of both CDNets are the same, with computational complexity of $\mathcal{O}({N_{\rm{c}}}{N_{\rm{t}}}^2+{N_{\rm{t}}}{N_{\rm{c}}}^2)$ and $\mathcal{O}({N_{\rm{t}}}{N_{\rm{c}}}S)$, respectively. The computational complexity of the LSTM part in RCDNet is $\mathcal{O}(S^2n)$, while the computational complexity of the transformer part in ACDNet is $\mathcal{O}(S^2n+Sn^2)$, including $\mathcal{O}(Sn^2)$ self-attention computation and $\mathcal{O}(S^2n)$ feed-forward computation. In summary, the computational complexity of RCDNet is $\mathcal{O}(S^2n+{N_{\rm{c}}}{N_{\rm{t}}}^2+{N_{\rm{t}}}{N_{\rm{c}}}^2+{N_{\rm{t}}}{N_{\rm{c}}}S)$. In comparison, the computational complexity of ACDNet is higher, $\mathcal{O}(S^2n+Sn^2+{N_{\rm{c}}}{N_{\rm{t}}}^2+{N_{\rm{t}}}{N_{\rm{c}}}^2+{N_{\rm{t}}}{N_{\rm{c}}}S)$, due to the square relationship between the complexity of the attention mechanism and the sequence length.

Another difference is the scope of information interaction. Recurrence computation brings a sequential receptive field. Each cell pays more attention to the information of the close time slots and tends to forget the details of the more distant time slots. As a result, RCDNet is more inclined to learn the common and stable features of all time slots, which makes it show good robustness in dealing with noise interference and error propagation. In comparison, with the attention mechanism that directly links all sequence data, ACDNet has a global receptive field on the past-present interaction, and thus, it can sensitively capture features without being limited by time order. As a result, especially in use cases under quasi-static mobility,  ACDNet often shows better performance since some past channels may have similar small-scale features to the present channel.

Comparisons of RCDNet and ACDNet are summarized in Table \ref{table_compare}. Despite the differences in characteristics, both RCDNet and ACDNet are implemented under the guidance of the CDNet architecture in Fig. \ref{CDNet_fig}, and thus, follow the same paradigm in both training and deployment, as presented in the next section.

\subsection{Training and Deployment of CDNets}\label{section3.4}
\subsubsection{Training Data}
Easy-to-collect training data, such as unlabeled data, can significantly reduce learning costs and facilitate widespread applications of DL schemes.
For CDNets, user channel sequences of arbitrary continuous time slots, $\{{{\bf{H}}_{t - m}}, \ldots, {{\bf{H}}_{t-1}}, {{\bf{H}}_{t}}\},~\forall m \ge 0$, can be used as the training data, since the inputs $\{{{\bf{H}}_{t - m}}, \ldots, {{\bf{H}}_{t-1}}, {{\bf{H}}_{t}^0}\}$ and output labels ${{\bf{H}}_{t}}$ can be generated directly from such channel sequences. Along with the communication process, such channel sequences naturally exist in large numbers in wireless systems. Thus, the available training data for CDNets is sizable, and collecting these data does not impose much extra burden on the system.

{\color{black}Meanwhile, there exists a valuable data augmentation method for CDNets' training set, i.e., generating more legitimate training data based on the collected training data to improve the quality of training. Since the channel samples of a short-term mobile sequence are in a finite neighborhood, we can swap the internal order or draw out certain sub-elements to get a new channel sequence whose samples are still in this neighborhood and thus can be regarded as a new channel sequence under with different moving speed or directions. Specifically, we extract several elements from $\{{{\bf{H}}_{t - m}}, \ldots, {{\bf{H}}_{t-1}}, {{\bf{H}}_{t}}\},~\forall m \ge 0$ to form a new sequence (the extracted elements can be repeatedly and do not need to remain the order in origin sequence) as the augmented data. In this way, even with only channel sequences in limited mobile modes, the network can see corresponding channel characteristics in more diverse mobile modes during training to improve the learned generalization. Therefore, this data augmentation method can significantly increase data utilization efficiency and reduce the data collection overhead required for training.}

\subsubsection{Loss Function and Optimizer}
To train the CDNets, mean square error (MSE) is used as the loss function, which can be written as
\begin{equation}
   {\rm{Loss}}\left( \Theta  \right) = \frac{1}{{num}}\sum\limits_{j = 1}^{num} {{{\left[ {\left\| {{{\bf{H}}_t} - {{\widehat {\bf{H}}}_t}} \right\|_2^2} \right]}_j}},
\label{MSE}
\end{equation}
where $\Theta$ is the parameter set of the CDNet, $j$ represents the $j$-th training data, {\color{black} $num$ is the total number of training data (including the augmented data) in the training set, and $\Vert \cdot \Vert_2$ is the Euclidean norm. In addition, $\Theta$ can be updated through the existing gradient descent-based optimizers, such as the adaptive momentum estimation (Adam) optimizer \cite{adam}. In this paper, the training of $\Theta$ is offline.}

\subsubsection{Deployment}
After CDNets have been trained, we obtain a usable $\rm{g}_{\rm{cd}}(\cdot)$ function that can deduce ${{\bf{H}}_{t}}$ based on ${{{\bf{H}}_{t - n}}, \ldots ,{{\bf{H}}_{t - 1}},{\bf{H}}_t^0}$. Meanwhile, this deduction process can be continued in an autoregressive manner, i.e., after the acquisition of ${{\bf{H}}_{t}}$, ${{\bf{H}}_{t}}$ can be engaged in the acquisition of ${{\bf{H}}_{t+1}}$, which can be written as
\begin{equation}
    {{\bf{H}}_{t + k}} = {{g}_{{\rm{cd}}}}\left( {{{\bf{H}}_{t + k - n}}, \ldots ,{{\bf{H}}_{t + k - 1}},{\bf{H}}_{t + k}^0} \right),~k = 0,1, \ldots.
\end{equation}
In this way, it is only necessary to use a high density of pilots to obtain ${{\bf{H}}_{0}}, {{\bf{H}}_{1}}, \ldots, {{\bf{H}}_{n-1}}$ in the first few time slots after accessing the system \cite{access_noma,access_che}, then ${{\bf{H}}_{n}}, {{\bf{H}}_{n+1}}, \ldots$ can be continuously obtained by estimating ${{\bf{H}}_{n}^0}, {{\bf{H}}_{n+1}^0}, \ldots$ based on only tiny pilots.

\section{Numerical Experiments}\label{section4}
In this section, we evaluate the performance of proposed RCDNet and ACDNet. We first introduce the experiment settings. Then, we show the experimental results, including accuracy of channel acquisition, convergence and generalization during training, relationship between deduction accuracy and number of available past channels, and robustness to lossy input. Finally, a specific example of serving a mobile user through trained CDNets is presented.

\subsection{Experiment Settings}\label{section4.1}
\subsubsection{Communication Scenario} 
\begin{figure}[htbp]
  \centering
  \includegraphics[width=0.85\linewidth]{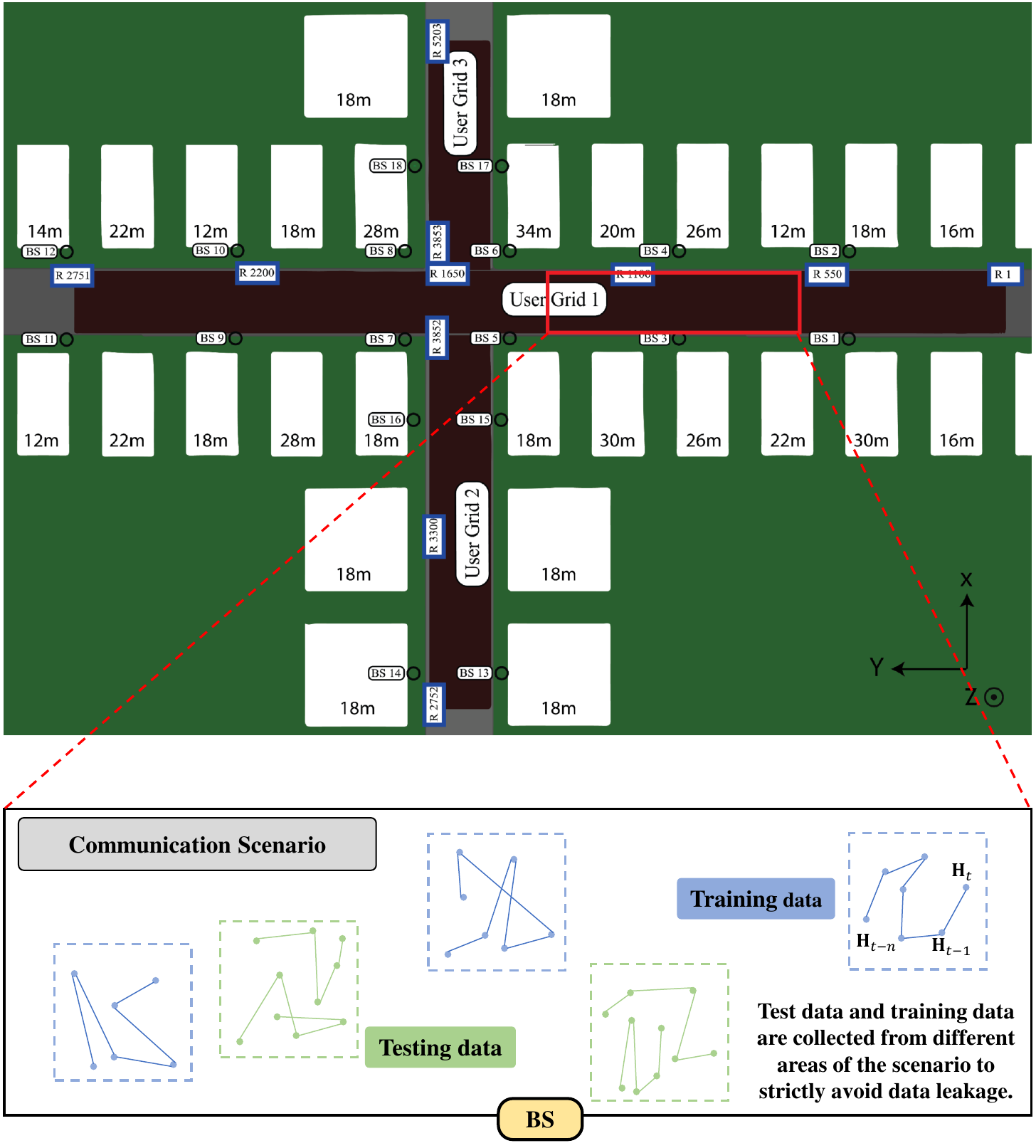}
  \caption{\small Using `O1' scenario in DeepMIMO dataset \cite{deepmimo} as experimental scenario, and collecting training and testing datasets from it.}
  \label{O1_fig}
\end{figure}

In this work, we use open-source raytracing-based DeepMIMO scenarios and datasets \cite{deepmimo} for experiments. Specifically, we use a typical outdoor scenario, ‘O1’, as the communication scenario. Fig. \ref{O1_fig}  shows the scattering environment of ‘O1’ scenario, a typical urban outdoor environment with two streets and one intersection. We set the BS 3 equipped with a ULA to serve the users in the red box area of Fig. \ref{O1_fig}. 

\begin{table}[htbp]\footnotesize
	\caption{\small Parameter settings for DeepMIMO datasets}
	\vspace{-0.8em}
	\begin{center}
		\begin{tabular}{ p{4cm}   p{4cm}}
			\toprule
			\textbf{Parameters} & \textbf{Value} \\
			\toprule
			Frequency band & 3.5GHz\\
			Bandwidth & 40MHz \\
			Base station & BS3 \\
			Antenna array form & ULA \\					
			Number of antennas ($N_\mathrm{t}$)  & 32     \\
            Number of subcarriers ($N_\mathrm{c}$) & 32     \\
            Number of paths  ($P$) & 25     \\  
            User area & R501 - R1400 \\
            Number of training data & 3000 \\
            Sequence length of each training data & 32 \\
            Number of mobile testing data & 20000 \\
            Number of quasi-static testing data & 20000 \\
            Sequence length of each testing data  & 17 (16 past channel and 1 present channel) \\
			\toprule
		\end{tabular}		
	\end{center}
	\vspace{-1.2em}
	\label{table_deepmimo}
\end{table}

In this scenario, we assume many users continuously communicate with the BS. The channel sequences in the communication process are collected as the training set and testing set. We collect the training and test data in different areas of the scenario to avoid data leakage, as shown in Fig. \ref{O1_fig}. Meanwhile, the user's motion pattern significantly affects the time correlation, which in turn affects the performance of approaches utilizing this correlation, including channel prediction and channel deduction. Therefore, we divide the testing set into two parts, one with channel sequences under high-speed motion and the other under quasi-static. The diverse testing sets will better demonstrate the relationship between performance and time correlation. As introduced in section \ref{section3.4}, the training set and the augmented data are used to train each channel acquisition scheme. Besides, in testing sets, the last channel of each channel sequence is the present channel, and the previous channels are past channels. The acquisition results of the present channel by each scheme will be compared with the true values to evaluate the performance. Table \ref{table_deepmimo} shows the detailed simulation settings for DeepMIMO datasets.

\begin{table}\footnotesize
    	\caption{\small \color{black} Parameter settings and FLOPs of CDNets and benchmarks}
	\vspace{-0.8em}
	\begin{center}
		\begin{tabular}{ p{4cm}   p{4cm}}
			\toprule
			\textbf{Parameters} & \textbf{Value} \\
			\toprule
			Batch size & 500\\
			Training epochs & 100000 \\
                Training steps & Training epochs × (3000 / 500) \\
                Learning rate & ${S^{ - 0.5}} \times {\rm{min}} ({step^{ - 0.5}},step \times {warmup\_step}^{ - 1.5})$\cite{transformer}, $S=512,~warmup\_step=4000$\\
			Loss function & MSE \\			
                Data Augmentation & Shown in Section \ref{section3.4} \\
                Structure parameters of RCDNet & $K_1=3,~K_2=3,~S=512$ \\
                Structure parameters of ACDNet & $K_1=3,~K_2=3,~K_3=6,~S=512$ \\
                \textcolor{black}{FLOPs of RCDNet} & \textcolor{black}{194.89 Million} \\ 
                \textcolor{black}{FLOPs of ACDNet} & \textcolor{black}{373.51 Million} \\ 
                \textcolor{black}{FLOPs of LSTM-based prediction} & \textcolor{black}{170.03 Million} \\ 
                \textcolor{black}{FLOPs of CMixer-based estimation} & \textcolor{black}{17.57 Million} \\ 
                Subset $\Omega$ of present channel to estimated &  $\allowdisplaybreaks \{{0,~l_{\rm{t}}, ~\ldots ,~(N_{\rm{t}}^0 - 1) \times l_{\rm{t}}} \} \otimes \{{0,~l_{\rm{c}}, ~\ldots ,~(N_{\rm{c}}^0 - 1) \times l_{\rm{c}}} \}$, $l_{\rm{t}} = {N_{\rm{t}}}//N_{\rm{t}}^0$, $l_{\rm{c}} = {N_{\rm{c}}}//N_{\rm{c}}^0$ \\
                Optimizer & Adam \cite{adam}   \\
			\toprule
		\end{tabular}		
	\end{center}
	\vspace{-2.2em}
	\label{table_training}
\end{table}

\subsubsection{Performance Indexes}
We use normalized MSE (NMSE) and cosine correlation $\rho$ \cite{csinet} between acquired channel and true channel as the performance indexes, which are defined as follows:
\begin{equation}
    \mathrm{NMSE} = \mathbb{E}\left\{ {\frac{{\left\| {{\mathbf{H}} - \widehat {\mathbf{H}}} \right\|_2^2}}{{\left\| {\mathbf{H}} \right\|_2^2}}} \right\},
\label{NMSE}
\end{equation}
and
\begin{equation}
    \rho {\rm{ = }}{\mathop{\mathbb{E} }\nolimits} \left\{ {\frac{1}{{{N_c}}}\sum\limits_{m = 1}^{{N_c}} {\frac{{|{{\widehat {\bf{h}}}_m}^H{{\bf{h}}_m}|}}{{||{{\widehat {\bf{h}}}_m}|{|_2}||{{\bf{h}}_m}|{|_2}}}} } \right\},
\label{rou}
\end{equation}
where ${{\bf{h}}_m}, {\widehat {\bf{h}}}_m$ are the original and acquired CSI of the $m$-th subcarrier, i.e. the $m$-th columons of CSI matrices $\mathbf{H}, \widehat {\mathbf{H}}$, respectively.

\subsubsection{Benchmarks and Training Settings}
{\color{black} For intuitive evaluation, we use two representative and competitive channel acquisition approaches as benchmarks to be compared. One is the channel estimation enhanced by CMixer-based channel mapping, and the CMixer contains 8 CMixer layers \cite{cmixer}. The other is the channel prediction based on an LSTM network \cite{lstm_pred1}, where the network structure is the same as the LSTM part in RCDNet. Details about these three schemes are shown in Section \ref{section1.2} and corresponding literature. Also, to ensure fair comparisons, we use the same training settings for all schemes, as shown in Table \ref{table_training}.  In addition, the floating operation points (FLOPs) of all schemes (the case of known present channel size is 4 antennas × 4 subcarriers) are also presented in Table \ref{table_training}. Moreover, the proposed DL methods are programmed based on the PyTorch library of version 1.12 \cite{pytorch}, and all simulations are implemented on an NVIDIA V100 server \cite{V100}. The training of RCDNet and ACDNet occupied about 4 GB and 7 GB of graphics memory, respectively.}

\subsection{Performance Evaluation} \label{section4.2}
\subsubsection{Accuracy of Acquired CSI}
\begin{figure*}[htbp]
\centering
  \includegraphics[width=0.8\textwidth]{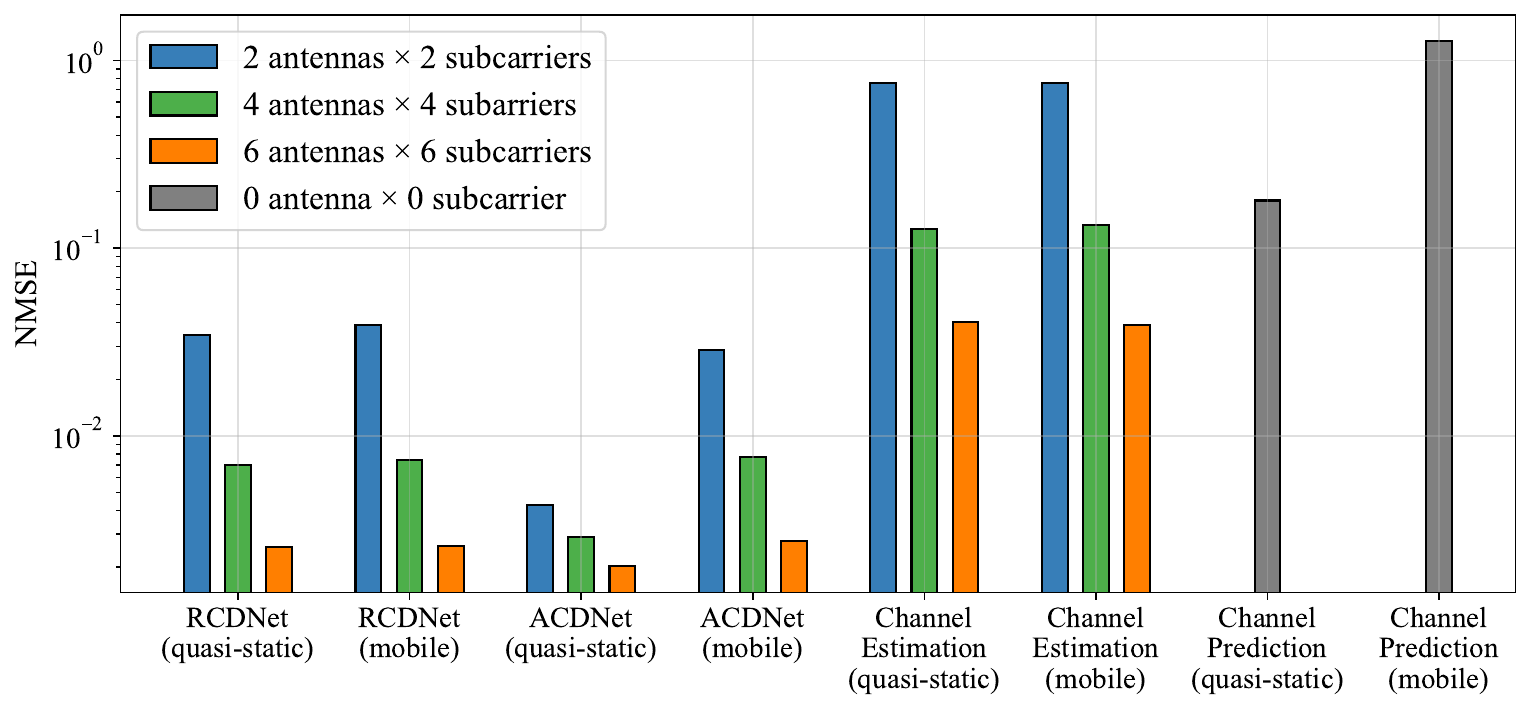}
      \caption{\small 
      {\color{black}NMSE of proposed CDNets and benchmarks under various estimated present channel sizes. The sizes of estimated present partial channel are shown in the legend and the size of whole channel is 32 antennas × 32 subcarriers.}
      }
     \vspace{0.4em}
  \label{nmse_fig}
\end{figure*}

\begin{table*}[]\footnotesize
\renewcommand{\arraystretch}{1.8}
\caption{\small \color{black} The cosine correlation $\rho$ of proposed CDNets and benchmarks under various estimated present channel sizes. The sizes of estimated present partial channel are shown in the leftest column and the size of whole channel is 32 antennas × 32 subcarriers.}
\vspace{-0.8em}
\begin{center}
\begin{tabular}{c|cccccc|cc}
\hline
\multirow{2}{*}{\makecell{Estimated channel size\\through pilot\\(antennas × subcarriers)}} & \multicolumn{2}{c|}{RCDNet}  & \multicolumn{2}{c|}{ACDNet}                         & \multicolumn{2}{c|}{Channel Estimation} & \multicolumn{2}{c}{Channel Prediction}    \\ \cline{2-9} 
 & \multicolumn{1}{c|}{quasi-static} & \multicolumn{1}{c|}{~~~mobile~~~~} & \multicolumn{1}{c|}{quasi-static} & \multicolumn{1}{c|}{~~~mobile~~~~} & \multicolumn{1}{c|}{quasi-static} & \multicolumn{1}{c|}{~~~mobile~~~~} & \multicolumn{1}{c|}{quasi-static} & ~~~mobile~~~~ \\ \hline
2 × 2 & \multicolumn{1}{c|}{0.9886} & \multicolumn{1}{c|}{0.9869} & \multicolumn{1}{c|}{0.9981} & \multicolumn{1}{c|}{0.9895} & \multicolumn{2}{c|}{0.5851} & \multicolumn{2}{c}{\multirow{3}{*}{\makecell{0.9416~~~~~~~~~0.7824\\(No pilot is used)}}}  \\ \cline{1-7}
4 × 4 & \multicolumn{1}{c|}{0.9973} & \multicolumn{1}{c|}{0.9972} & \multicolumn{1}{c|}{0.9988} & \multicolumn{1}{c|}{0.9969} & \multicolumn{2}{c|}{0.9348}    \\ \cline{1-7}
6 × 6 & \multicolumn{1}{c|}{0.9989} & \multicolumn{1}{c|}{0.9989} & \multicolumn{1}{c|}{0.9991} & \multicolumn{1}{c|}{0.9988} & \multicolumn{2}{c|}{0.9797}     \\ \hline
\end{tabular}
\end{center}
\vspace{-0.5em}
\label{table_rou}
\end{table*}

\begin{figure}[htbp]
\centering
  \includegraphics[width=0.45\textwidth]{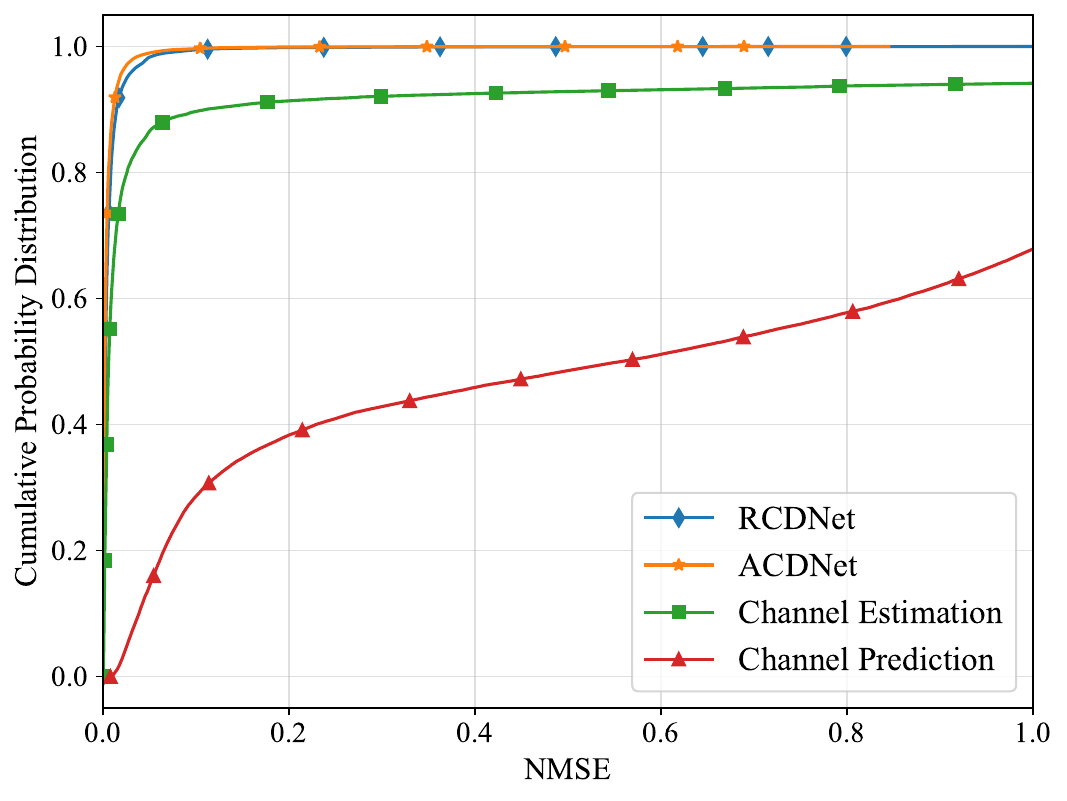}
  \vspace{-0.3em}
      \caption{\small \color{black} Cumulative probability distribution of errors (NMSE) between the acquired channel and the true channel, counted all tested samples including quasi-static and mobile cases. }
     \vspace{-0.3em}
  \label{cdf_fig}
\end{figure}

The accuracy of the acquired CSI is a crucial performance evaluation indicator. Fig. \ref{nmse_fig} and Table \ref{table_rou} show the mean NMSE and $\rho$ over the testing sets under various sizes of known present channels, respectively. The proposed RCDNet and ACDNet always provide lower NMSE and higher $\rho$ compared to benchmarks. Specifically, the size of the known present channel is smaller, the performance gain of CDNets over channel estimation is larger, to 13 dB in NMSE and 40 percentage points in $\rho$ for the case where only 2 antennas × 2 subcarriers are known. Moreover, compared to channel prediction, even if only 2 antennas × 2 subcarriers of the present channel is additionally known, the performance gain of the proposed CDNets reaches 7dB in NMSE and 4 percentage points in $\rho$ under the quasi-static use case, and 16dB in NMSE and 20 percentage points in $\rho$ under the mobile case.  
In summary, proposed channel deduction approaches are able to provide high-quality acquisition even under various complex mobile patterns and tiny known present subchannels, solving the weakness of estimation and prediction. 
Meanwhile, for the internal comparisons of CDNets, ACDNet outperforms RCDNet under the quasi-static case, reflecting the superiority of attention mechanism over recurrence computation in the sensitivity of detailed feature extraction, which has been analyzed in Section \ref{section3.4}.

\begin{figure*}[htbp]
\centering
  \includegraphics[width=0.96\textwidth]{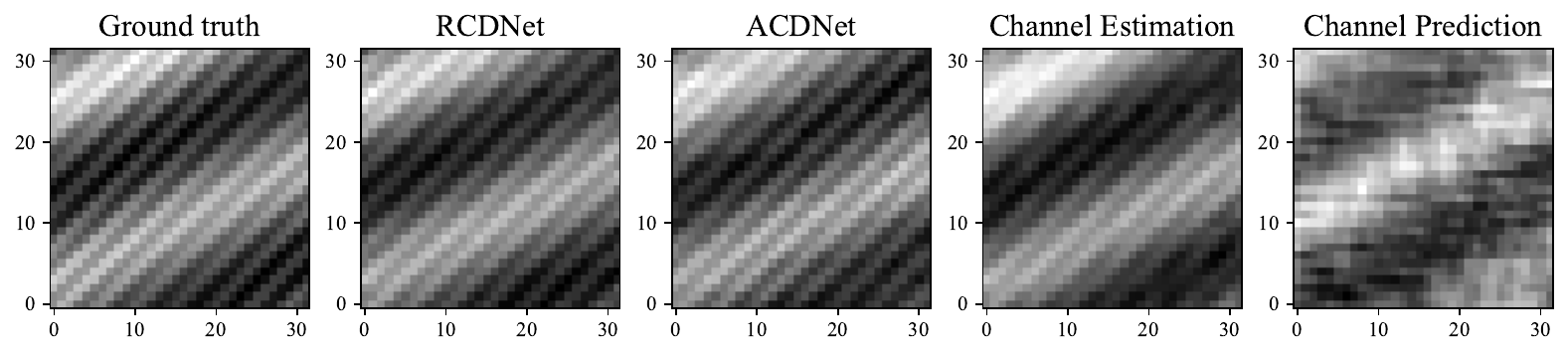}
      \caption{\small Grayscale visualization of the true CSI and acquired CSI. {\color{black}(the vertical axis is the antenna domain, and the horizontal axis is the frequency domain)}}
     \vspace{-0.8em}
  \label{visualization_fig}
\end{figure*}

In addition, Fig. \ref{cdf_fig} shows the cumulative probability distribution of errors of proposed CDNets and benchmarks. Here, we use NMSE as the error metric and set known present channel size as 4 antennas × 4 subcarriers for an example, and the rest of the experiments in this paper all follow these settings. The channel acquisition errors (NMSE) on all test data of the RCDNet and ACDNet are tiny and concentrated within 0.1, which means that it can provide a high-quality channel acquisition service for almost all users very stably. Besides, Fig. \ref{visualization_fig} shows the grayscale visualization of the acquired and true channels on randomly chosen test data. It can be intuitively seen that CDNets characterize the present channel with high accuracy. In contrast, channel estimation loses some detail features due to limited available information, and channel prediction only provides approximation limited by mobile uncertainty.

\subsubsection{Convergence and Generalization during Training}
\begin{figure}[htbp]
\centering
  \includegraphics[width=0.44\textwidth]{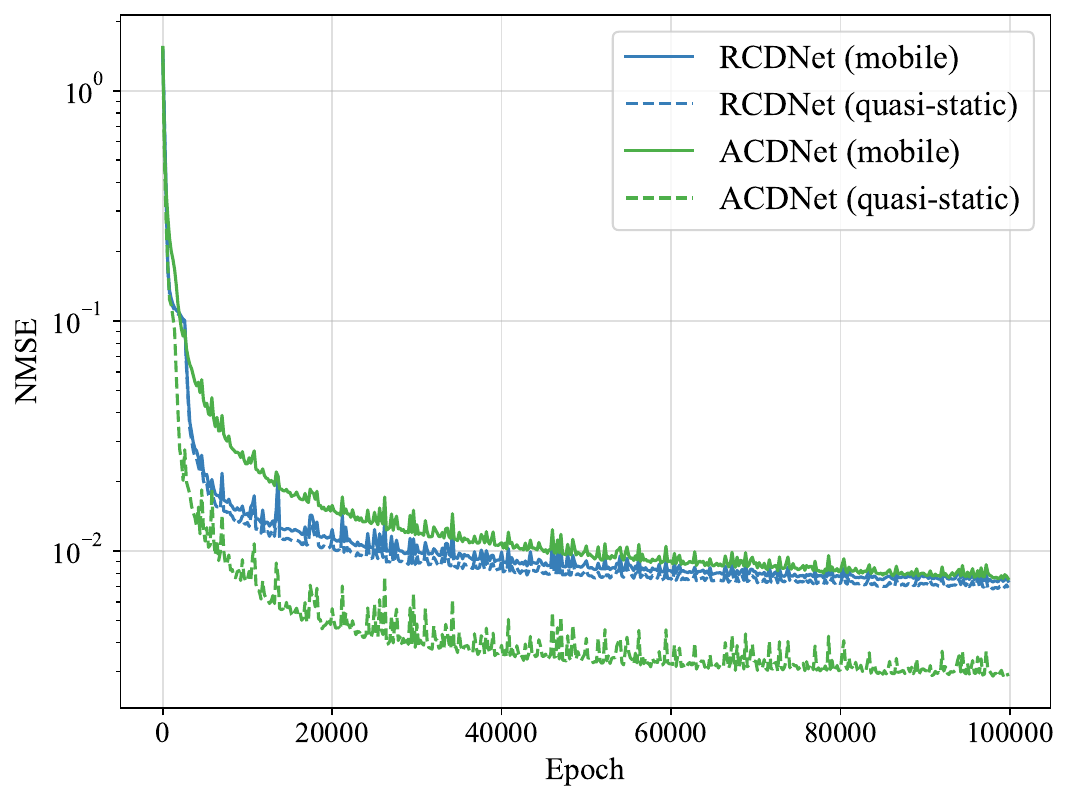}
      \caption{\small The generalization of CDNets on testing sets (NMSE) during training.}
     \vspace{-0.8em}
  \label{convergence_fig}
\end{figure}

The convergence of neural networks is a common concern in DL-enabled implementations. Stable convergence allows the algorithm to be quickly and widely applied and indicates that the designed learning structure suits the target task. Fig. \ref{convergence_fig} shows the mean NMSE of RCDNet and ACDNet over the testing sets during training. Despite the slight fluctuation of NMSE caused by stochastic gradient descent, the CDNets maintain a generally stable convergence during the training process, with a progressive improvement in generalization on both quasi-static and mobile cases. Moreover, CDNets can achieve excellent performance even after only short-term training at the beginning, which makes it quite promising for online learning and fine-tuning where fast training is required.

\subsubsection{Acquired Accuracy versus Number of Past Channels}
\begin{figure}[htbp]
\centering
  \includegraphics[width=0.44\textwidth]{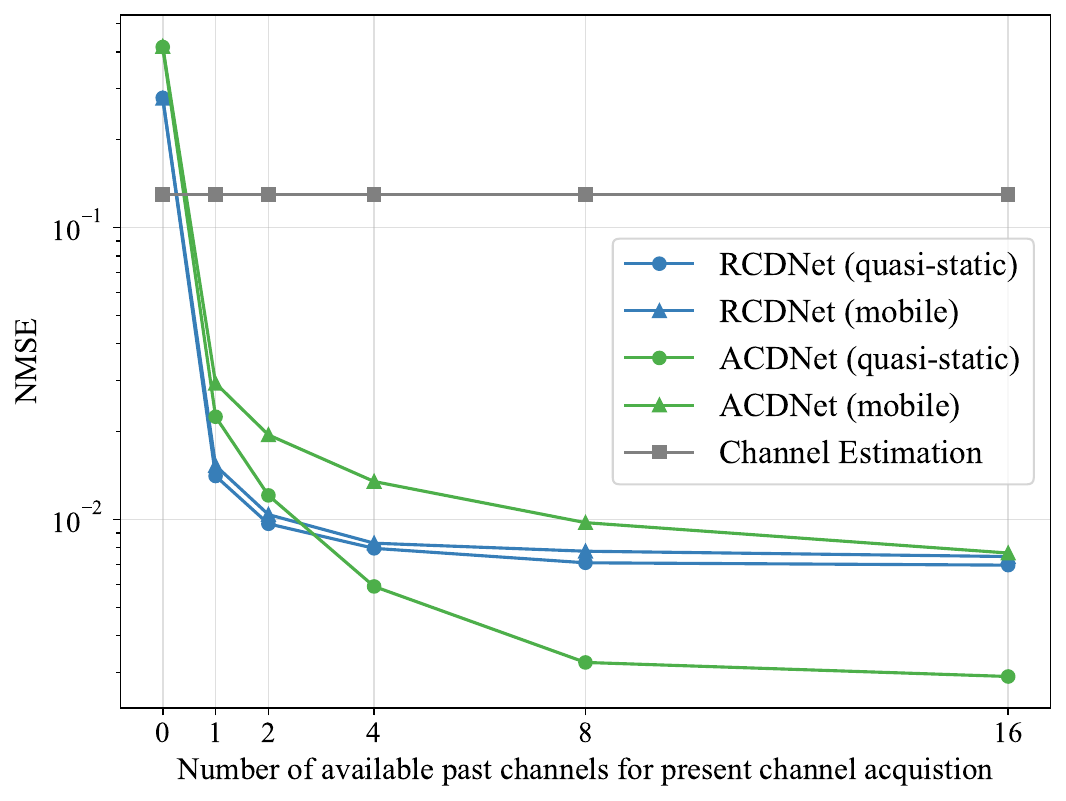}
      \caption{\small NMSE versus number of available past channels. (The size of estimated partial channel through pilots is 4 antennas × 4 subcarriers) }
     \vspace{-0.5em}
  \label{pastlen_fig}
\end{figure}

The performance gains of channel deduction over channel estimation derive from extracting additional usable information from past channels. This part evaluates the relationship between the performance of CDNets and the number of past channels, as shown in Fig. \ref{pastlen_fig}. In both RCDNet and ACDNet, NMSE decreases and then nearly saturates as the number of available past channels increases. This phenomenon indicates that CDNets effectively obtain usable features from past channels to improve performance and obtain more adequate features as available past channels become more. Even with only one available past channel, CDNets can provide significant performance gains of more than 7 dB over channel estimation. In addition, comparing RCDNet and ACDNet, RCDNet performs better when the number of past channels is small ($n\le4$) since a small amount of recurrence does not tend to result in information forgetting, and the RNN can achieve sufficient information accumulation. In contrast, ACDNet is more advantageous than RCDNet when the number of past channels is large ($n>4$) since the global receptive field of attention mechanism can better utilize the information in long sequences.

\subsubsection{Robustness of CDNets}

\begin{figure}
	\centering
	\subfigure[NMSE under adding different disturbance $\sigma$ to past and known partial present channel.]{\label{noise_fig}
		\includegraphics[width=0.44\textwidth]{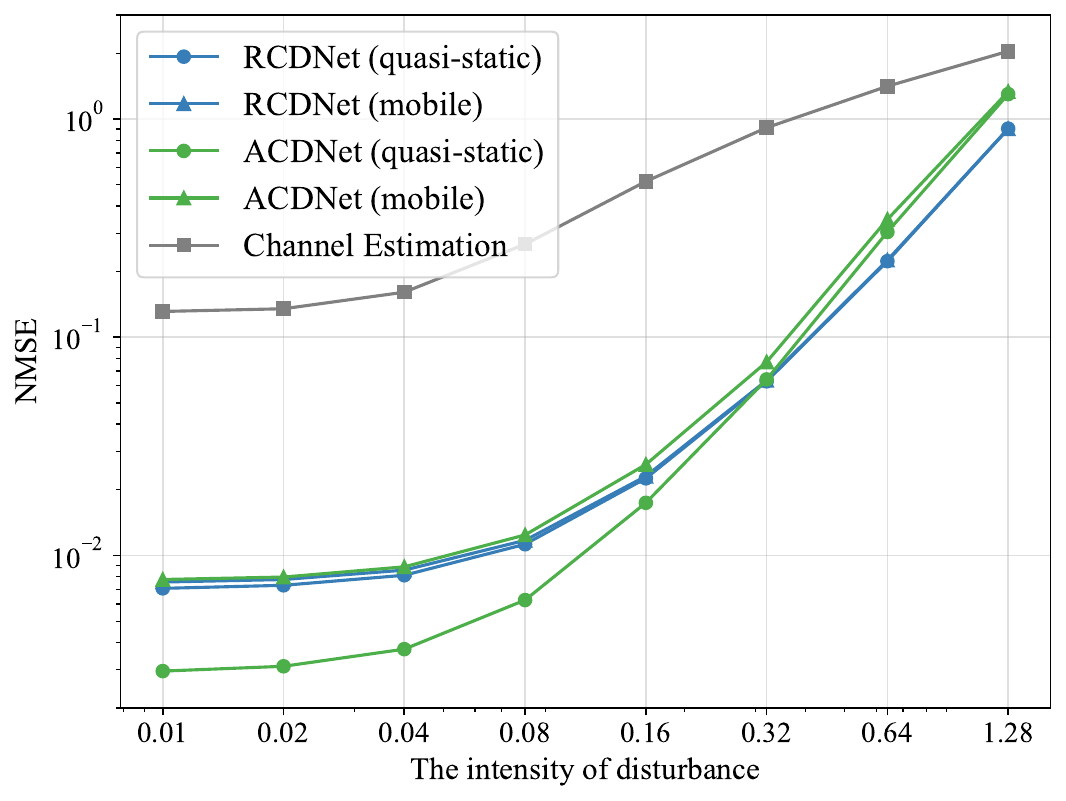}
	}
	\centering
	\subfigure[NMSE under only adding different disturbance $\sigma$ to past channel.] {\label{noise_past_fig}
		\includegraphics[width=0.44\textwidth]{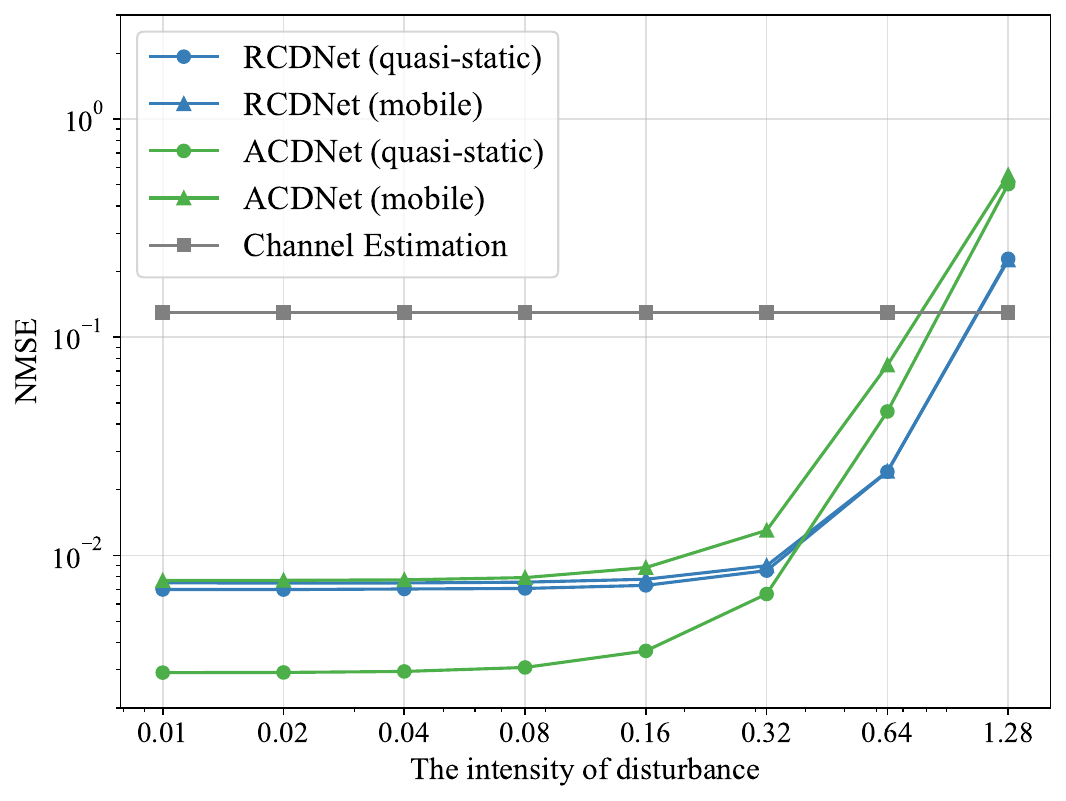}
	}
	\caption{\small \color{black} Robustness evaluation of proposed CDNets. (The size of estimated partial channel through pilots is 4 antennas × 4 subcarriers)
 }
	\vspace{-0.5em}
\label{robustness_fig}
\end{figure}

\begin{figure}[htbp]
  \centering
  \begin{minipage}[l]{0.01\textwidth}
    ~
  \end{minipage}
  \begin{minipage}[l]{0.455\textwidth}
    \centering
    \includegraphics[width=\textwidth]{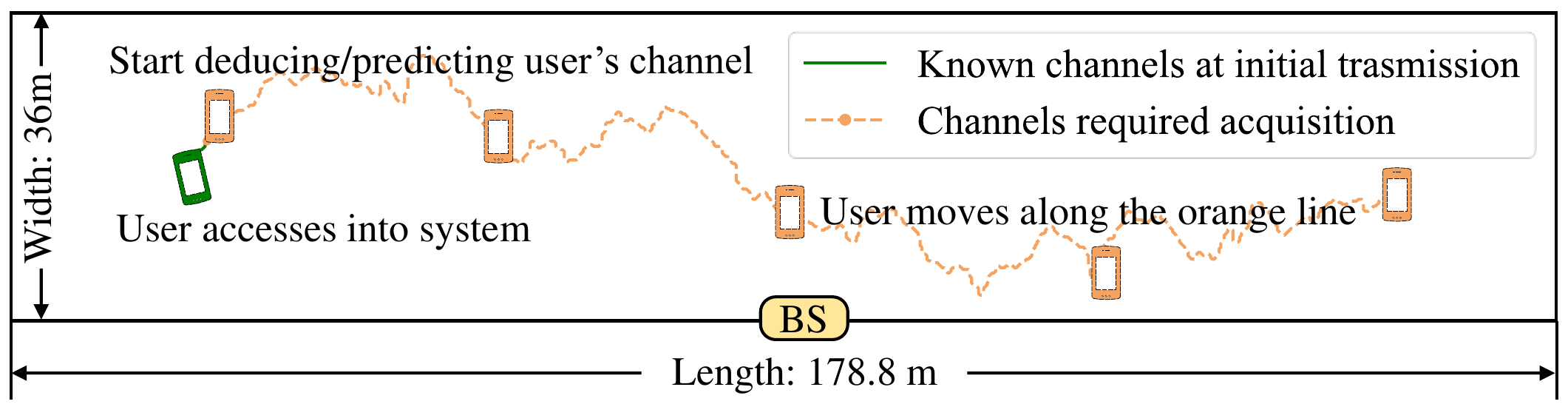}
    \end{minipage}
    \begin{minipage}[t]{0.455\textwidth}
    \caption*{\footnotesize (a) The trajectory of a user's movement in `O1' scenario. During this movement (2024 time slots), autoregressively acquiring the user channel through approaches including deduction and prediction that rely on time correlation.}
\end{minipage}
  \\[0.2em]
  \begin{minipage}[t]{0.49\textwidth}
    \centering
  \includegraphics[width=\textwidth]{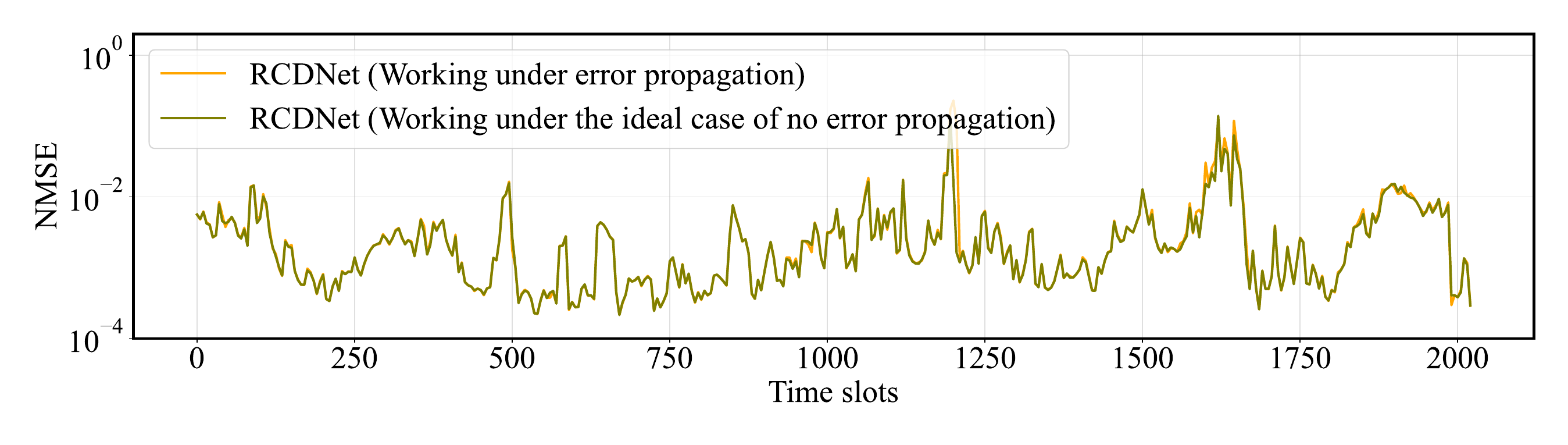}
  \end{minipage}
  \\[-0.4em]
  \begin{minipage}[t]{0.49\textwidth}
    \centering
  \includegraphics[width=\textwidth]{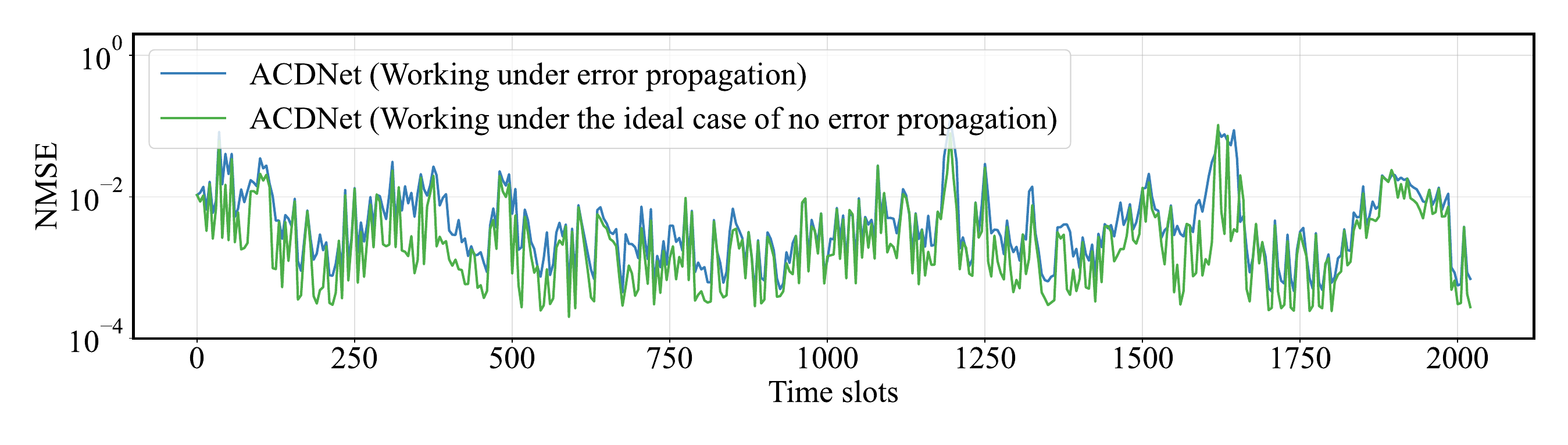}
  \end{minipage}
  \\[-0.4em]
  \begin{minipage}[t]{0.49\textwidth}
    \centering
  \includegraphics[width=\textwidth]{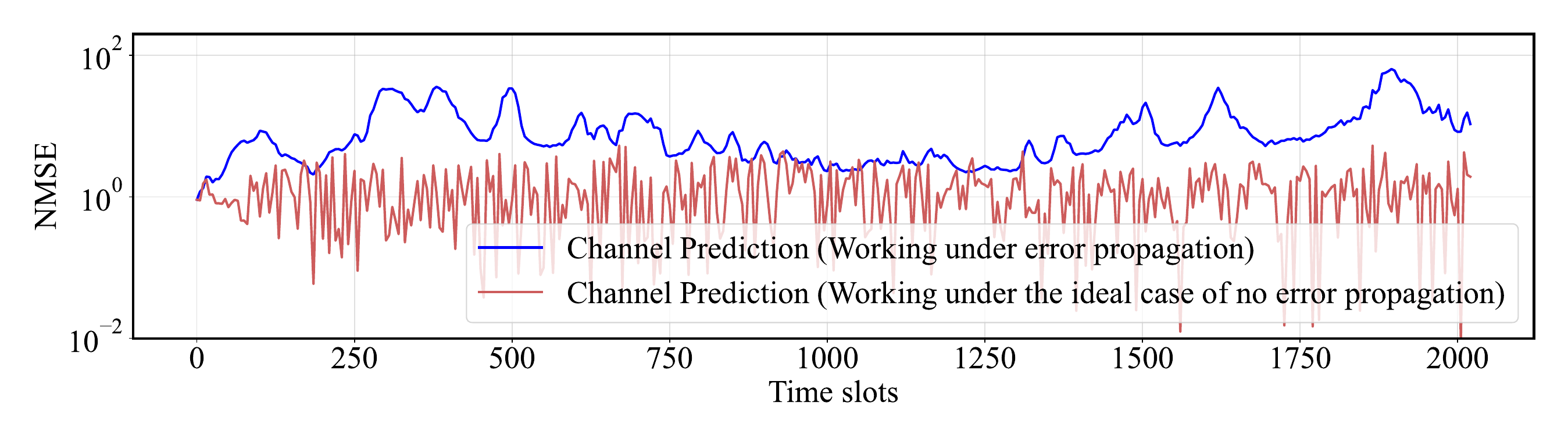}
  \\[-0.3em]
  \end{minipage}
  \begin{minipage}[t]{0.49\textwidth}
\vspace{-0.6em}
    \caption*{\footnotesize (b) NMSE between the acquired channel and the true channel during movement.}
    \vspace{-0.5em}
\end{minipage}
  \caption{\small \color{black}Continuously Serve a Mobile User through CDNets.}
  \vspace{-1em}
\label{application_fig}
\end{figure}

In practical applications, the known information used to acquire the channel is often lossy due to noise, interference, and errors, which challenges the robustness of the channel acquisition algorithm. This subsection evaluates the robustness of CDNets, including two typical cases. One is to evaluate the performance of CDNets when all inputs $\{{{{\bf{H}}_{t - n}}, \ldots ,{{\bf{H}}_{t - 1}},{\bf{H}}_t^0}\}$ are lossy, which is important for the network to mitigate noise and interference. The other is to evaluate the performance of CDNets when the past channels $\{{{\bf{H}}_{t - n}}, \ldots ,{{\bf{H}}_{t - 1}}\}$ are lossy and the known present subchannel ${\bf{H}}_t^0$ is ideal, in order to test the ability of leveraging new information to self-calibrate to cope with error propagation. Unlike the other subsections that suppose ideal known CSI, in this subsection, we add disturbance with different intensities to the ideal CSI to simulate lossy CSI. Specifically, we add disturbance by the following way \cite{fdma_positioning}: ${{\bf{H}}_{{\rm{dis}}}} = {\bf{H}} \odot {\bf{D}}$, where ${{\bf{H}}_{{\rm{dis}}}}$ is the disturbed CSI matrix, ${\bf{H}}$ is the ideal CSI matrix, $\odot$ is the Hadamard product, ${\bf{D}}$ is the disturbed matrix and each element of ${\bf{D}}$ is an independent and identically distributed Gaussian random variable obeying $N\left( {1,{\sigma ^2}} \right)$. $\sigma$ simulates the deviation degree of the lossy channel from the ideal channel.

Fig. \ref{robustness_fig}(a) illustrates the relationship between the performance of CDNets and the deviation degree $\sigma$ in the case where the past channels and the present subchannel are lossy. CDNets show better robustness than channel estimation in resisting damage to known information. In addition, at low deviation $(\sigma\le0.16)$, ACDNet maintains its performance advantage over RCDNet, while at high deviation $(\sigma>0.16)$, RCDNet is more robust than ACDNet since its learning structure is more inclined to learn stable common features, which has been analyzed in Section \ref{section3.3}. Moreover, Fig. \ref{robustness_fig}(b) illustrates the relationship between the performance of CDNets and the deviation degree $\sigma$ in the case where the past channels are lossy and the present subchannel is ideal. Only in the case where the past channels are so lossy $(\sigma\ge1.28)$ that the necessary transmission of the past time slots cannot be guaranteed, CDNets do not perform as well as the estimation methods that do not rely on the past information. In all other cases, CDNets significantly outperform channel estimation. In addition, compared to the lossy present subchannel case in Fig.~\ref{robustness_fig}(a), the advantage of RCDNet over ACDNet in robustness is more significant when the present subchannel is ideal. This is thanks to RCDNet's cumulative computational structure, where the sequential receptive field pays particular attention to the present information and thus better utilizes the present information for calibration.

\subsubsection{Continuously Serve a Mobile User through CDNets} 

Thanks to the autoregressive continuous deployment and excellent robustness to error propagation, the performance gains of CDNets are not only on specific time slots but also throughout the continuous communication process. In this section, we use trained CDNets to continuously provide channel acquisition service to a mobile user, visually displaying the proposed approach's application value. Fig. \ref{application_fig}(a) shows a mobile user and its mobile trajectory. The user equipment acquires complete CSI with the high-density pilots in the initial $n$ time slots ($n$ is set as 8), and the trajectory during these time slots is shown as the green line. Then, for the subsequent 2024 time slots, the user moves (and may also be stationary at some time slots), and its trajectory is shown as the orange line. The CDNets continuously provide channel deduction services based on the known CSI of previous $n$ time slots and a partial estimate (4 antennas × 4 subcarriers) of present channel.

{\color{black} Fig. \ref{application_fig}(b) presents the specific performance of CDNets in these 2024 time slots. During continuous 2024 time slots, the CDNets are able to stably acquire channels with high quality. Specifically, by comparing with the deduction based on the ideal past channel, it can be found that even in the autoregressive process with error propagation, both RCDNet and ACDNet are only slightly performance-damaged. Meanwhile, RCDNet exhibits a lower performance degradation against error propagation than ACDNet, thanks to its cumulative learning structure, which has been detailedly analyzed in Section \ref{section3.3}. In addition, we also apply channel prediction to this autoregressive acquisition process. 
Even if prediction is based on the ideal past channels, it still suffers from significant performance fluctuations due to difficulty coping with irregular user movements. Moreover, when based on the acquired lossy past channels instead of the ideal past channels, the prediction approach suffers more severe performance degradation due to error propagation. } 
The comparison between the two approaches also reflects that the proposed channel deduction successfully solves the pain points of prediction and has promising application value.


\section{Conclusion} \label{conclusion}
In this paper, we propose the channel deduction approach for channel acquisition in MIMO-OFDM systems. By means of DL techniques, we design two specific implementations, ACDNet and RCDNet, and provide the related data collection and augmentation, training, and deployment methods. Numerical results illustrate the effectiveness and superiority of CDNets, including high-accuracy channel acquisition with tiny pilot cost, excellent robustness to lossy inputs, and successful continuous tracking of user channels under error propagation and complicated mobility. These capacities are significant for addressing the urgent needs of mobile channel acquisition in MIMO-OFDM systems, demonstrating the promising application value of the channel deduction approach.

For the high-dimensional channel acquisition task, we first delve into what are the necessary features for channel representation and which low-cost information medium can provide them. We also notice the strong technical support that DL can provide in freely fusing implicit features from multiple mediums and representing high-dimensional data. This collaboration of ideas and techniques guides the construction and implementation of the channel deduction approach. In this work, DL techniques are not used to improve algorithms on existing wireless tasks but to empower the generation of more foundational tasks and technical tools based on wireless systems' requirements and physical mechanisms. We hope that our work can provide help and inspiration for the application of MIMO and the development of wireless AI.

{\color{black}
\vspace{1em}
\appendix
\vspace{0.3em}
\section{Additional Experiment Results under Uniform Plane Array System}
In the appendix, we provide additional experimental results under a uniform planar array (UPA) system, as a supplement to the results under the ULA system presented in the main text, in order to broaden the scope of the evaluation. In this UPA system, the BS employs an 8×4 planar array, while all other settings remain consistent with those in the main text. We evaluate the performance of the proposed CDNets and benchmarks in this new system, as shown in Fig. \ref{nmse_fig_UPA} and Table \ref{table_rou_UPA}. In this UPA system, the proposed CD method still achieves high-precision channel acquisition with tiny pilot overhead, and it still offers significant performance gains compared to the benchmarks.

\begin{figure*}[t]
\centering
  \includegraphics[width=0.8\textwidth]{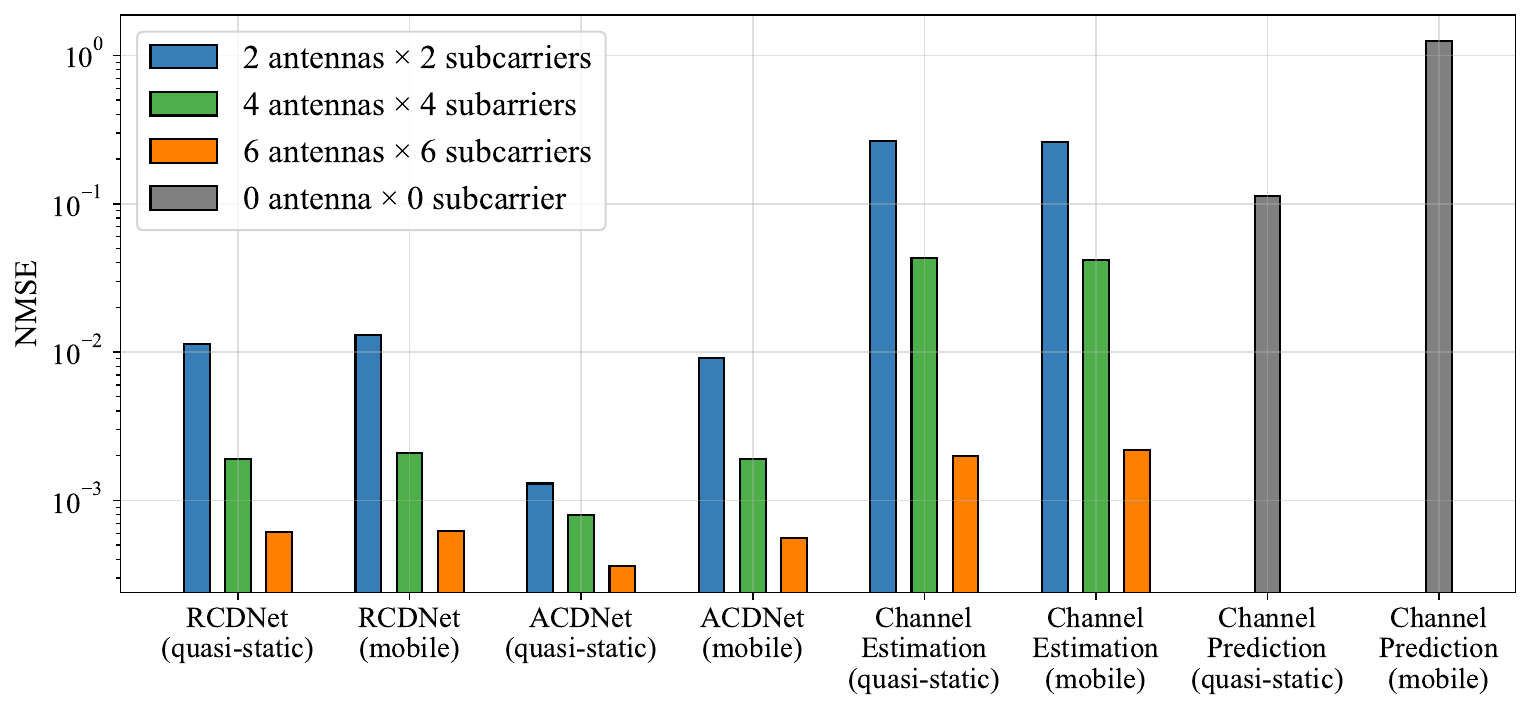}
      \caption{\small \color{black}
      NMSE of proposed CDNets and benchmarks under various estimated present channel sizes (Under a UPA system). The sizes of estimated present partial channel are shown in the legend and the size of whole channel is 32 antennas × 32 subcarriers.}
     \vspace{0.4em}
  \label{nmse_fig_UPA}
\end{figure*}

\begin{table*}[htbp]\footnotesize
\renewcommand{\arraystretch}{1.8}
\caption{\small \color{black} The cosine correlation $\rho$ of proposed CDNets and benchmarks under various estimated present channel sizes (Under a UPA system). The sizes of estimated present partial channel are shown in the leftest column and the size of whole channel is 32 antennas × 32 subcarriers.}
\vspace{-0.8em}
\begin{center}
\begin{tabular}{c|cccccc|cc}
\hline
\multirow{2}{*}{\makecell{Estimated channel size\\through pilot\\(antennas × subcarriers)}} & \multicolumn{2}{c|}{RCDNet}  & \multicolumn{2}{c|}{ACDNet}                         & \multicolumn{2}{c|}{Channel Estimation} & \multicolumn{2}{c}{Channel Prediction}    \\ \cline{2-9} 
 & \multicolumn{1}{c|}{quasi-static} & \multicolumn{1}{c|}{~~~mobile~~~~} & \multicolumn{1}{c|}{quasi-static} & \multicolumn{1}{c|}{~~~mobile~~~~} & \multicolumn{1}{c|}{quasi-static} & \multicolumn{1}{c|}{~~~mobile~~~~} & \multicolumn{1}{c|}{quasi-static} & ~~~mobile~~~~ \\ \hline
2 × 2 & \multicolumn{1}{c|}{0.9965} & \multicolumn{1}{c|}{0.9961} & \multicolumn{1}{c|}{0.9995} & \multicolumn{1}{c|}{0.9970} & \multicolumn{2}{c|}{0.9035} & \multicolumn{2}{c}{\multirow{3}{*}{\makecell{0.9760~~~~~~~~~0.8572\\(No pilot is used)}}}  \\ \cline{1-7}
4 × 4 & \multicolumn{1}{c|}{0.9993} & \multicolumn{1}{c|}{0.9992} & \multicolumn{1}{c|}{0.9996} & \multicolumn{1}{c|}{0.9992} & \multicolumn{2}{c|}{0.9786}    \\ \cline{1-7}
6 × 6 & \multicolumn{1}{c|}{0.9997} & \multicolumn{1}{c|}{0.9997} & \multicolumn{1}{c|}{0.9998} & \multicolumn{1}{c|}{0.9997} & \multicolumn{2}{c|}{0.9989}     \\ \hline
\end{tabular}
\end{center}
\vspace{-0.5em}
\label{table_rou_UPA}
\end{table*}
}

\bibliographystyle{IEEEbib}
\bibliography{ref}

\end{document}